\documentclass[acmsmall,nonacm]{acmart}

\usepackage{soul} 
\usepackage{graphicx}
\usepackage{titlesec}
\usepackage{multirow}
\usepackage{cleveref}

\usepackage{tikz} 
\usetikzlibrary{arrows.meta} 
\usetikzlibrary{decorations.markings}
\usetikzlibrary{backgrounds}
\usetikzlibrary{positioning,chains,fit,shapes,calc}
\usetikzlibrary{graphs,quotes,trees}
\usepackage{subcaption}
\tikzset{font={\fontsize{8pt}{10}\selectfont}}

\usepackage{dsfont} 

\AtBeginDocument{%
  \providecommand\BibTeX{{%
    \normalfont B\kern-0.5em{\scshape i\kern-0.25em b}\kern-0.8em\TeX}}}

\begin{document}

\title{Private Graph Data Release: A Survey}

%
\author{Yang Li}
\email{kelvin.li@anu.edu.au}
\author{Michael Purcell}
\email{michael.purcell1@anu.edu.au}
\affiliation{%
  \institution{School of Computing, The Australian National University}
}
\author{Thierry Rakotoarivelo}
\email{thierry.rakotoarivelo@data61.csiro.au}
\author{David Smith}
\email{david.smith@data61.csiro.au}
\author{Thilina Ranbaduge}
\email{thilina.ranbaduge@data61.csiro.au}
\affiliation{%
  \institution{Data61, CSIRO}
}
\author{Kee Siong Ng}
\email{keesiong.ng@anu.edu.au}
\affiliation{%
  \institution{School of Computing, The Australian National University}
}


\def\M{\mathcal{M}}
\def\Sm{\mathcal{S}}
\def\Om{\mathcal{O}}
\def\e{\varepsilon}
\def\Prob{\mathrm{Pr}}
\def\X{\mathcal{X}}
\def\Lap{\mathrm{Lap}}
\definecolor{trgreen}{RGB}{28,172,0}

\newcommand{\tr}[1]{{\color{black}#1}}
\newcommand{\ds}[1]{{\color{black}#1}}
\newcommand{\kl}[1]{{\color{black}#1}}
\newcommand{\thierry}[1]{{\color{black}#1}}
\newcommand{\mpp}[1]{{\color{black}#1}}

\begin{abstract}
The application of graph analytics to various domains has yielded tremendous societal and economical benefits in recent years.
However, the increasingly widespread adoption of graph analytics comes with a commensurate increase in the need to protect private information in graph data, especially in light of the many privacy breaches in real-world graph data that was supposed to preserve sensitive information.
This paper provides a comprehensive survey of private graph data release algorithms that seek to achieve the fine balance between privacy and utility, with a specific focus on provably private mechanisms.
Many of these mechanisms are natural extensions of the Differential Privacy framework to graph data, but we also investigate more general privacy formulations like Pufferfish Privacy that address some of the limitations of Differential Privacy.
We also provide a wide-ranging survey of the applications of private graph
data release mechanisms to social networks, finance, supply
chain, and health care.
This survey paper and the taxonomy it provides should benefit practitioners and researchers alike in the increasingly important area of private analytics and data release.
\end{abstract}

\maketitle

\section{Introduction}
\label{sec:intro}

Graph analytics refer to the methods and tools used to study and understand the
relationships that exist between vertices and edges within and between graphs\footnote{Throughout, these terms are used interchangeably, graph and network, node and vertex, edge and link.}~\cite{nisar2013}.
In contrast to traditional analytics on tabular data, graph analytics is
concerned with the flows, the structure, and the relationships between
the vertices and the edges that compose graph data. Examples of common
statistics of interests for such graph data include node degrees and the related
degree distribution, centrality metrics (e.g., degree, closeness, betweenness, etc.), subgraph counts (e.g., triangle, k-star, etc), and various distance metrics (e.g., diameter, eccentricity, etc.)~\cite{sanfeliu83}.

The applications of graph analytics to various domains has yielded tremendous societal and economical
benefits. Indeed, they are used to understand and track the spread of diseases
within communities; to study proteins, chemical configurations, and interactions
in the design of novel medicines; to uncover irregular fraud patterns in financial
records; and increase the resilience of supply chains in various
sectors such as agrifood, healthcare, and manufacturing~\cite{pourhabibi2020,tan2019,leskovec2007}.
As in traditional data analytics, these benefits are even greater when graph
analytics are applied to collections of graph data from various stakeholders, e.g., uncovering international money laundering by analyzing graph data from different
financial institutions across countries~\cite{weber2018a}. In \cite{mam2019,met2020}, the authors estimate global value of graph analytics to be about USD 600 million in 2019. They further project that this value could grow to USD 2 to 3 billion by 2026.

One critical challenge in having graph data available across different parties is
{\em trust}. The parties making graph data available need to trust that
confidential and private information within such data will be protected, whereas the
analysts using such data need to trust that they still contain the information
required for their needs.
Such trust is challenging to achieve and maintain. Indeed, there have been many cases where the release of graph data has lead to the inadvertent exposure of sensitive personal information.
For example, a recent study shows that
specific graph analytics applied to an easily available social network graph
could predict with good accuracy the sexual orientation of individuals~\cite{jernigan2009}. Further works have shown how attacks
can be constructed to extract sensitive information from graphs that have been treated with some form of anonymization, including identity disclosure and attribute disclosure \cite{ji2017,beigi2020survey,backstrom2007wherefore,casas2017survey}.

Several authors have studied the fine balance between privacy and utility for various types of data~\cite{brickell2008,Sankar2013,li2009}. These mechanisms are often categorized as either
{\em non-provable} or {\em provable}~\cite{wu2010survey,ji2017}. Provable privacy mechanisms offer mathematical guarantees about the privacy protection or the utility that they can provide. Non-provable privacy mechanisms do not provide such strong theoretical guarantees and thus are more empirical in nature. 
Many of these provable methods are based on a formal definition of privacy known as
Differential Privacy (DP) \cite{dwork2014algorithmic}. Differential Privacy is a
property of a release mechanism, i.e., an algorithm that can be used to publish information about a data set.
A differentially private release mechanism essentially guarantees that its outputs
(e.g., a synthetic data set or a query answer) will be almost indistinguishable from the output it would produce if any one individual's data was removed from the data set. As such, differentially private release mechanisms provide a degree of
\emph{plausible deniability} for every individual whose data may be included in the data.

Some surveys have studied the category of non-provable privacy mechanisms
for graph data. For example, \citep{wu2010survey} catalogs several non-provable
methods into 3 classes, namely k-anonymity, probabilistic, and
generalization. Similarly, \citep{casas2017survey} describes
non-provable approaches for the release of entire graphs,
thus not considering the release of graph statistics or metrics only.
More recent surveys have included some descriptions of
provable privacy mechanisms for graph data. For example, ~\cite{ji2017}, briefly 
describes
the emerging DP-based mechanisms for graph
data at the time and recognizes that DP-based methods for graph data were still in their infancy. Another study~\cite{abawajy2016}
provides a more thorough survey of some DP-based graph mechanisms, including
edge DP and node DP methods. However, it focuses solely on DP-based mechanisms that
were proposed in the context of social network graphs and does not discuss
other domain applications. A recent preprint
\cite{jiang2020}
also describes DP-based mechanisms only for social network graphs.
Finally, a recently published survey on privacy preserving data publishing~\cite{majeed2021} also has a dedicated section on graph data. It also considers only the application domain of social networks. Furthermore, it does not distinguish between
provable and non-provable approaches and provides only a brief overview of DP-based
mechanisms and no mention of other provable (but non-DP-based) methods.
Thus, to the best of our knowledge, there has not yet been a thorough survey of
provable privacy mechanisms for graph data, both DP and non-DP, and which considers a wider set of
domain applications.
This paper fills that gap.
It provides a comprehensive study of the published
state-of-the-art methods and algorithms used to provide provable privacy guarantees for graph analytics in a variety of application domains.

To guide our survey of privacy mechanisms for graph data, we propose a tree-based
classification of the papers that we review in this paper.
At a high level, this classification differentiates between mechanisms that release
an entire privacy-enhanced graph and others that release only statistics or query responses
on given graph data. We then further differentiate between non-provable and provable mechanisms.
The goal of this proposed taxonomy is to allow data custodians to easily navigate 
the large list of existing mechanisms and identify the mechanisms which could be fit-for-purpose for their needs.
We acknowledge that other taxonomies could have been equally adopted,
and that some of these alternatives may be more useful for  other readers of this survey (e.g., scientists
looking for knowledge gaps in the area). 
Thus, we also discuss another alternative classification of some existing works along
different application domains.

The remainder of this paper is organized as follows.
\Cref{sec:method} presents our methodology. It introduces the taxonomy used for the surveyed contributions, describes the different
related categories in that taxonomy, and describes the criteria for a given work to be assigned to a
specific category. 
\Cref{sec:stagraph} focuses on mechanisms 
that return 
responses to specific statistical queries on graph data. 
In contrast, \Cref{sec:syngraph} focuses on mechanisms that return entire graphs. 
While Sections~\ref{sec:stagraph} and~\ref{sec:syngraph} specifically
discuss non-provable and DP-based provable mechanisms, \Cref{sec:limits}
introduces other provable privacy definitions, which address some of the shortcomings of
DP in the context of graph data.
\Cref{sec:app} provides an overview of the different domain applications where
the previously surveyed contributions may be applied and some example use cases.
\Cref{sec:discuss} discusses some existing empirical studies on private graph analytics,
and some common limitations shared by several of the surveyed
contributions, which suggest future research opportunities.
\Cref{sec:end} concludes this survey.

\section{Method and Background}
\label{sec:method}

We present the taxonomy that we use to differentiate between the different
classes of privacy mechanisms that we surveyed. We then review the background concepts that underlie differential privacy (DP) and formally define
DP in the context of graph data.
We conclude this section by describing some of the fundamental differentially private release mechanisms upon which many other mechanisms are based.

\subsection{Method}
%


%
%
%

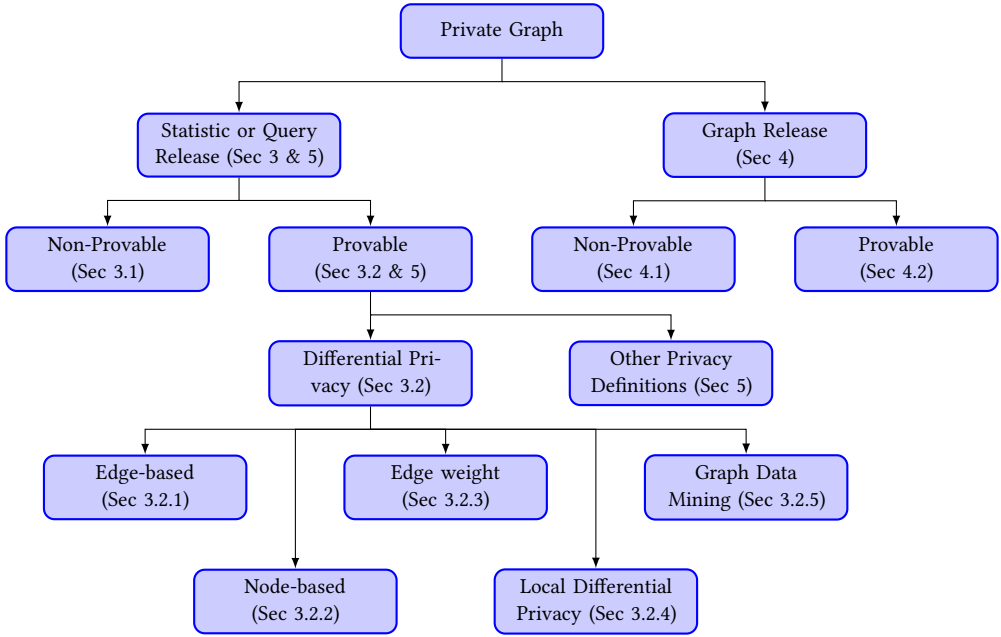
\begin{figure}
    \centering
    \begin{tikzpicture}[%
auto,
block/.style={
	rectangle,
	draw=blue,
	thick,
	fill=blue!20,
	text width=7em,
	align=center,
	rounded corners,
	minimum height=2em
}
]
\draw (-0.25,0) node[block] (privGra) {Private Graph};

\path (-3.75,-1.5) node[block] (statsRe) {Statistic or Query Release (Sec 3 \& 5)}
(3.25,-1.5) node[block] (graRe) {Graph Release\\ (Sec 4)}

(1.5,-3) node[block] (nonProv1) {Non-Provable\\ (Sec 4.1)}
(5,-3) node[block] (prov1) {Provable \\(Sec 4.2)}

(-5.5,-3) node[block] (nonProv2) {Non-Provable\\ (Sec 3.1)}
(-2,-3) node[block] (prov2) {Provable\\ (Sec 3.2 \& 5)}

(2,-4.5) node[block] (othPriv) {Other Privacy Definitions (Sec 5)}
	(-2,-4.5) node[block] (dp) {Differential Privacy (Sec 3.2)}
	
	(-5,-6) node[block] (edgeBased) {Edge-based\\ (Sec 3.2.1)}
	(-1,-6) node[block] (edgeWeight) {Edge weight\\ (Sec 3.2.3)}
	(3,-6) node[block] (graDM) {Graph Data Mining (Sec 3.2.5)}
	(-3,-7.5) node[block] (nodeBased) {Node-based\\ (Sec 3.2.2)}
	(1,-7.5) node[block] (ldp) {Local Differential Privacy (Sec 3.2.4)}
	;

\draw (privGra.south) -- ++(0,-0.3) coordinate (p11);
\draw (p11) -- ++(0,0) coordinate (p12);
\draw[-latex] (p12) -| (statsRe.north);
\draw[-latex] (p12) -| (graRe.north);

\draw (graRe.south) -- ++(0,-0.3) coordinate (p21);
\draw (p21) -- ++(0,0) coordinate (p22);
\draw[-latex] (p22) -| (nonProv1.north);
\draw[-latex] (p22) -| (prov1.north);

\draw (statsRe.south) -- ++(0,-0.3) coordinate (p31);
\draw (p31) -- ++(0,0) coordinate (p32);
\draw[-latex] (p32) -| (nonProv2.north);
\draw[-latex] (p32) -| (prov2.north);

\draw (prov2.south) -- ++(0,-0.3) coordinate (p41);
\draw (p41) -- ++(0,0) coordinate (p42);
\draw[-latex] (p42) -| (othPriv.north);
\draw[-latex] (p42) -| (dp.north);

\draw (dp.south) -- ++(0,-0.3) coordinate (p51);
\draw (p51) -- ++(0,0) coordinate (p52);
\draw[-latex] (p52) -| (edgeBased.north);
\draw[-latex] (p52) -| (nodeBased.north);
\draw[-latex] (p52) -| (edgeWeight.north);
\draw[-latex] (p52) -| (ldp.north);
\draw[-latex] (p52) -| (graDM.north);

\end{tikzpicture}
    \caption{Classification of the surveyed research on private graph data release.}
    \label{fig:struct}
\end{figure}

\Cref{fig:struct} presents the structure that we used to classify existing
contributions on privacy-preserving graph analytics. It also links
each part of this classification to the corresponding sections of this paper.
We decided on this structure
%
%
to provide a clear delineation of the contributions of this survey and identify the key focus areas of this survey. \Cref{tab:taxo} provides an overview of the surveyed published contributions following our proposed taxonomy.

We begin with a top-level differentiation between {\em graph release} mechanisms which release an entire (transformed graph), and {\em query release} mechanisms which release (transformed) responses to specific queries about a graph. We then progress to more specific areas as we traverse down the taxonomy.

As we traverse downwards, we differentiate between various notions of privacy. At the second level of our taxonomy we differentiate between provable and non-provable privacy. We then restrict our attention to provable privacy and  differentiate between differentially private query release mechanisms, and query release mechanisms that are based on other provable notions of privacy (e.g., Pufferfish Privacy). We concentrate most of our attention on differentially private mechanisms due to DP's widespread use and acceptance in the research community. Finally, at the bottom level of our taxonomy we differentiate between the various notions of graph differential privacy. 
   
        

Notice that \Cref{fig:struct} suggests that the primary focus of this paper is on provably private mechanisms that release statistics about graph data. Non-provably private mechanisms are discussed for completeness and to provide context for our discussion of provable mechanisms. For more details of non-provably private release of graph data the interested reader is referred to  \cite{casas2017survey, wu2010survey}.


\begin{table}[!t]
\renewcommand{\arraystretch}{1.2}
\vspace{-5mm}
\caption{Classification of the surveyed papers on private graph data release}
\label{tab:taxo}
\vspace{-2mm}
\footnotesize
\centering
\begin{tabular}{|l|l|l|l|l|} 
\hline
\multirow{17}{*}{\begin{tabular}[c]{@{}l@{}}Graph\\Statistics\\or Query\\Release\end{tabular}} & \multicolumn{3}{l|}{Non-Provable} & \cite{backstrom2007wherefore,korolova2008link,narayanan2009anonymizing} \\ 
\cline{2-5}
 & \multirow{16}{*}{Provable} & \multirow{13}{*}{\begin{tabular}[c]{@{}l@{}}Differential\\Privacy (DP)\\based\end{tabular}} & \multirow{4}{*}{Edge DP} 
  
 & {\begin{tabular}[c]{@{}l@{}}Outlink \cite{task2012guide}, Clustering Coefficient \cite{wang2013learning},\\
 Eigenvectors \cite{wang2013differential,ahmed2019publishing}, Graph Clustering \cite{mulle2015privacy},\\
 Community Detection \cite{nguyen2016detecting}, Edge Weight \cite{costea2013qualitative,li2017differential}\\
 Egocentric Betweenness Centrality \cite{roohi2019differentially}\end{tabular}}\\ 
\cline{5-5}
 &  &  &  & Subgraph Counting: \cite{nissim2007smooth,karwa2011private,lu2014exponential,zhang2015private,chen2013recursive,proserpio2014calibrating} \\ 
\cline{5-5}
 &  &  &  & Degree Sequence: \cite{proserpio2012workflow,hay2009accurate,karwa2012differentially} \\ 
\cline{5-5}
 &  &  &  & Cut Query: \cite{gupta2012iterative,blocki2012johnson,Upadhyay2013} \\ 
\cline{4-5}
 &  &  & \multirow{3}{*}{Node DP} 
 & {\begin{tabular}[c]{@{}l@{}} Erd\H{o}s-R\'{e}nyi Model Parameter \cite{borgs2015private,borgs2018revealing,sealfon2019efficiently},\\
 Difference Sequence \cite{song2018differentially}\end{tabular}}\\ 
\cline{5-5}
 &  &  &  & Degree sequence: \cite{kasiviswanathan2013analyzing,day2016publishing,raskhodnikova2016lipschitz} \\ 
\cline{5-5}
 &  &  &  & Subgraph Counting: \cite{blocki2013differentially,kasiviswanathan2013analyzing,ding2018privacy} \\ \cline{4-5}
 &  &  & Edge Weight DP & \cite{sealfon2016shortest} \\
\cline{4-5}
 &  &  & Local DP & \cite{sun2019analyzing,ye2020} \\ 
\cline{4-5}
 &  &  & \multirow{4}{*}{Graph Mining} & Frequent Pattern Mining: \cite{shen2013mining,xu2016differentially}\\ 
\cline{5-5}
 &  &  &  & Subgraph Discovery: \cite{Kearns16} \\ 
\cline{5-5}
 &  &  &  & Clustering: \cite{pinot2018} \\ 
\cline{5-5}
 &  &  &  & Graph Embedding: \cite{xu2018,zhang2019,raskhodnikova2016lipschitz}  \\
 \cline{5-5}
 &  &  &  & Graph Neural Networks:
 \cite{de2021,mueller2022,igamberdiev2021,sajadmanesh2021,papernot2016}  \\ 
\cline{3-5}
 &  & \multirow{3}{*}{Beyond DP} & Correlated Data DP & \cite{kifer2011no,kifer14,liuMC16,srivatsa2012,almadhoun2020,zhao2017} \\ 
\cline{4-5}
 &  &  & Pufferfish Privacy & \cite{KiferM12,kifer14,HeMD13,yang2015bayesian,song2017} \\ 
\cline{4-5}
 &  &  & Others & \cite{ghosh2017,rastogi2009relationship,gehrke2011towards} \\ 
\hline
\hline
\multirow{5}{*}{\begin{tabular}[c]{@{}l@{}}Graph\\Release\end{tabular}} & \multicolumn{3}{l|}{Non-Provable} & \cite{sweeney2002k,samarati2001protecting,chakrabarti2006,wu2010survey,casas2017survey,kiranmayi2020review} \\ 
\cline{2-5}
 & \multicolumn{2}{l|}{\multirow{4}{*}{Provable}} & Generative Models & \cite{sala2011sharing,mahadevan2006,hay2009accurate,wang2013preserving,iftikhar2020dk,mir2012differentially,leskovec2007scalable,gleich2012moment,xiao2014differentially,wang2013differential,jorgensen2016publishing,zhang2015private} \\ 
\cline{4-5}
 & \multicolumn{2}{l|}{} & Graph Matrix &  \cite{wang2013differential,chen2014correlated,xiao2014differentially,brunet2016novel,blum2005practical,hardt2011beating,kapralov2013differentially,blocki2012johnson,Upadhyay2013} \\ 
\cline{4-5}
 & \multicolumn{2}{l|}{} & Local DP & \cite{qin2017,gao2018local} \\ 
\cline{4-5}
 & \multicolumn{2}{l|}{} & Iterative Refinement & \cite{gupta2012iterative,proserpio2012workflow,proserpio2014calibrating} \\
\hline
\end{tabular}
\end{table}

\subsection{Background}

\begin{table*}[!t]
\renewcommand{\arraystretch}{1.2}
\vspace{-5mm}
\caption{\label{tab:note}Notation used in this survey}
\label{t:notations}
\vspace{-2mm}
\centering
\setlength\tabcolsep{2pt}
\ds{\small{
\begin{tabular}{l l l l l}
\hline
$||.||_1$ & $L_1$ norm && $\mathcal{X}$&Data universe \\ 
$\M$ & Randomized algorithm&&$\mathds{N}$& Natural numbers\\
$\Om$ & Output from a randomized algorithm&&$\mathbb{R}^n$& Real numbers of dimension $n$\\
$D^n$&Data set domain of dimension $n$&&$D$ & Data set\\
$G$&Graph&&$V$ & Graph vertices\\
$E$ & Graph edges&& $M_E$&Edge Matrix \\ 
$S(G)$&Degree sequence of $G$&&$d(\cdot)$&Degree of $(\cdot)$\\
$\Bar{d}$&Average degree&&$\e$ & Privacy budget\\
$\delta$ & Privacy budget approximation&&$q(\cdot)$& query function\\
$GS_f$&Global sensitivity of $f$&&$LS_f$&Local sensitivity of $f$\\
$RS_f$&Restricted sensitivity of $f$&&$SS_f$&Smooth sensitivity of $f$ \\
$\Lap(\lambda)$&Laplace mechanism, scale factor $\lambda$&&$\mu$&Data projection \\
$\Delta(g)$&$L_1$ sensitivity of $g(\cdot)$ && $\Delta q$&Range sensitivity of $q(\cdot)$\\
$\theta$&Degree threshold&&$x \oplus y$ & Symmetric difference between sets $x$ and $y$\\
$\omega$&Graph weight function&&&\\
\hline
\end{tabular}}}
\end{table*}

\Cref{tab:note} introduces the notation that we will use in this survey. Following the terminologies in \cite{dwork2014algorithmic}, given a universe $\X$ of $n$ distinct data values, we consider a data set $D \in \mathds{N}^{|\X|}$ as a length-$n$ vector of counts, where $D_i$ is the number of times the $i$-th element in $\X$ occurs in the data set.
The $L_1$ norm of the data set is defined as 
\begin{equation*}
    ||D||_1 = \sum_{i=1}^n D_i.
\end{equation*}
The distance between two data sets $D^{(1)}, D^{(2)} \in \mathds{N}^{|\X|}$ is then defined as 
\begin{equation*}
    ||D^{(1)}-D^{(2)}||_1 = \sum_{i=1}^n |D^{(1)}_i-D^{(2)}_i|,
\end{equation*}
the total count differences between the two data sets. We say two data sets are \textit{neighboring}, denoted by $D^{(1)} \sim D^{(2)}$, if they differ on at most one coordinate (or record),\footnote{Some versions of neighboring data sets allow a record in a data set $x$ to be replaced by a different value to obtain another data set $y$. This implies a distance 2 if distance is measured by $L_1$-norm, but distance 1 if it is measured by edit distance.} 
i.e., 
\begin{equation}
\label{equation:nbr database}
    ||D^{(1)}-D^{(2)}||_1 \le 1.
\end{equation}
The classical definition of differential privacy \cite{dwork2014algorithmic} is then as follows:
\begin{definition}
\label{def:dp}
A randomized algorithm, $\M $, 
guarantees $(\e,\delta)$\textbf{-differential privacy}, 
if for any two neighboring data sets $D^{(1)}$ and $D^{(2)}$, and 
any subset of outputs $\Om\subseteq \it{range}(\M)$, we have
\begin{align}
\Prob{[\M(D^{(1)}) \in \Om]} \leq \exp(\e){\Prob{[\M(D^{(2)}) \in \Om]}}+\delta.
\end{align}
\end{definition}
We refer to $\e > 0$ as the \emph{privacy budget},
with smaller values of $\e$ providing
stronger privacy protection. When $\delta = 0$, we sometimes say that we have pure differential privacy. When $\delta>0$, we sometimes say that we have approximate differential privacy. Furthermore, when $\delta > 0$, its value is typically less than the inverse of the number of records in the data set. This precludes the (blatantly non-private) mechanism that simply returns a random record in response to a query $\M$.

Differential Privacy as defined in \Cref{def:dp} is largely a \emph{syntactic} construct. 
The semantics of \Cref{def:dp}, in terms of the indistinguishability of the prior and posterior probabilities after seeing the result returned by a Differential Privacy mechanism, can be found in \cite{kasiviswanathan14}.

For graph data, differential privacy can be expressed as follows. 
Suppose we have records from a universe $\mathcal{X}$,
a graph $G \in 2^{\mathcal{X} \times \mathcal{X}}$ is one where the nodes are records from $\mathcal{X}$. 

\begin{definition}
\label{def:dp for graph}
A randomized algorithm $\mathcal{M}$ with domain $2^{\mathcal{X} \times \mathcal{X}}$ is $(\epsilon,\delta)$\textbf{-differentially private} for a distance function $d$ if for any subset $\Om \subseteq {\it range}({\mathcal M})$ and $G_1, G_2 \in 2^{\mathcal{X}\times\mathcal{X}}$ such that $d(G_1,G_2) \leq k$:
\[ \Pr[\mathcal{M}(G_1) \in \Om] \leq \exp(\epsilon)\Pr[\mathcal{M}(G_2) \in \Om] + \delta, \]
where the probability is over the randomness in the mechanism $\mathcal{M}$.
\end{definition}

There are several forms of differential privacy that are particularly relevant in our context. These include node and edge differential privacy, both of which are described in detail in Section 3.2. These forms of differential privacy can be understood in terms of node and edge neighboring graphs. For node differential privacy, the distance function $d$ is the symmetric difference between the node sets of two graphs. For edge differential privacy, the distance function $d$ is the symmetric difference between the edge sets of the two graphs.


Importantly there are some simple differentially private mechanisms upon which many more complicated mechanisms are based. Chief among these is the {\em Laplace mechanism}~\cite{dwork2006calibrating,dwork2014algorithmic}. The Laplace mechanism works by adding noise drawn from the Laplace distribution to the result of a real-valued query. A fundamental result in differential privacy is that if we choose the scale for the Laplace noise appropriately, then the Laplace mechanism preserves  $(\varepsilon, 0)$ differential privacy.

We denote a random variable drawn from a Laplace (symmetric  exponential) distribution with mean 0 and scale $\lambda$ (or equivalently a variance $\sigma^2=2\lambda^2$) as $Y \sim \Lap(\lambda)$. Recall that the Laplace distribution has the following probability density function:
\begin{equation*}
    f(x|\lambda) = \frac{1}{2\lambda} \exp{\left(-\frac{|x|}{\lambda}\right)}.
\end{equation*}
The scale of the noise used in the Laplace mechanism depends on both the value of the privacy budget $\varepsilon$ and on the {\em global sensitivity} of the underlying query. The global sensitivity of a query is a measure of the largest possible amount of change in a function when one record is removed from a data set. More precisely, if we let $\Delta f$ denote the global sensitivity of a query $f$ then we have
\begin{equation}
\label{equation:global sensitivity}
    \Delta f = \max_{D^{(1)} \sim D^{(2)}} ||f(D^{(1)}) - f(D^{(2)})||_1.
\end{equation}
To ensure that the Laplace mechanism preserves $(\varepsilon, 0)$ differential privacy, then we must choose $\lambda > \Delta f/\varepsilon$.
The global sensitivity therefore is related to how much noise is required in the worst case to protect the privacy of an individual record in the data set.
 
There are other statistical-distribution based mechanisms used to implement differential privacy. 
The exponential mechanism~\cite{mcsherry2007} is a particularly important one-sided mechanism for producing differentially private graphs, particularly with respect to categorical answers. It uses a function $q(D, \mathcal{O})$ to represent how good an output $\mathcal{O}$ is for a data set $D$ (or equivalently a graph $G$). The exponential mechanism is the natural building block for answering queries with arbitrary utilities (and arbitrary non-numeric
range), while preserving differential privacy \cite{dwork2014algorithmic}. Given some arbitrary
range $R$, the exponential mechanism is defined with respect to some
utility function $q : \mathds{N}^{|\mathcal{X} |} \times \mathbb{R} \rightarrow \mathbb{R}^n$, which maps data set and output pairs
to utility scores. Intuitively, for a fixed data set (e.g., a graph $G$), the user prefers that
the mechanism outputs some element of $R$ with the maximum possible utility score.

As with the Laplace mechanism, the precise shape of the distribution from which outputs are drawn depends on the sensitivity of a function. In the case of the Exponential mechanism, the relevant quantity is the sensitivity of the utility function with respect to its data set argument. More precisely, the relevant quantity is
\begin{equation}
    \Delta q = \max_{r \in \mathbb{R}}\max_{x,y,\|x-y\|_1 \leq 1} |q(x,r)-q(y,r)|.
\end{equation}
Notice $\Delta q$ does not describe the sensitivity of $q$ with respect to its range argument. That is, we are interested in how the utility of each output changes for neighboring data sets but not in how the utility of each data set changes for neighboring outputs. To ensure that the exponential mechanism preserves $(\varepsilon, 0)$ differential privacy, we need the probability mass of each output to be proportional to $\exp\left(\varepsilon q(x,r) / 2\Delta q\right)$.


The application of these fundamental differentially private mechanisms in the context of graph release will be described in the following sections.


\section{Private Release of Graph Statistics or Queries}
\label{sec:stagraph}

There are two different approaches to protecting sensitive information from an analyst who can ask questions about a particular graph. The first approach is to act as a proxy for the analyst by querying the graph on their behalf. Then, the exact query responses can be transformed in some privacy-preserving way before being passed to the analyst. The second approach is to release a synthetic graph that is a close approximation to the true one but is guaranteed to be private according to some privacy framework.  The analyst can use the synthetic graph to compute answers to their queries. This section surveys works of the first type. The second type will be surveyed in \Cref{sec:syngraph}. 

We refer to methods of the first type as private query release mechanisms. These mechanisms typically take a graph, a class of graph queries, and some additional privacy parameters as inputs. They return privacy-preserving responses to the specified graph queries as outputs. Frequently, they do this by computing non-private query responses on the underlying graph and then transforming those responses in some way to ensure that they are privacy preserving. Although these mechanisms are query dependent, their advantages are at least two-fold. 
First, they explicitly define what information about the graph will be provided to the analyst. This makes it possible to identify which graph features need to be considered in the design of these release mechanisms and which do not.
Second, the noise needed to protect against a known class of queries is in general much less than that required to protect against the much larger class of all possible queries that an analyst could ask about a synthetic graph. As a general rule, adding less noise results in better query utility. 


\Cref{subsec:nonprovable graph query} briefly addresses the class of non-provable mechanisms
for privately releasing graph statistics,
while \Cref{subsec:provable graph query} extensively discusses the class of provable mechanisms, in particular those that are based on DP.




\subsection{Non-Provable Private Release of Graph Statistics}
\label{subsec:nonprovable graph query}


There is a reasonably large literature on statistical disclosure control \cite{willenborg01, hundepool2012} for releasing statistics on tabular data, including techniques like value generalization, cell suppression \cite{cox80}, micro-aggregation \cite{fayyoumiO10}, randomization \cite{gouweleeuw98}, and anonymization \cite{liuT08}. 
Most of these techniques have been applied to graph data, but their na{\"i}ve application is often inefficient or ineffective.

In~\cite{backstrom2007wherefore}, in the context of social networks where the curator replaces
names with meaningless unique identifiers, both active and passive attacks are presented and shown to be effective in revealing the existence of edges between users. Some of these attacks can be carried out using a single anonymized copy of the network and require relatively little effort from the attacker.
Similar work has been done by \citet{korolova2008link}, which shows an attacker recovering a significant fraction of sensitive edges through link analysis and getting a good picture of the whole network. 
\citet{narayanan2009anonymizing} also propose an algorithm to re-identify anonymized participants that were represented as graph vertices and apply it to social networks from Twitter and Flickr.
%
All these results show that mathematically rigorous definitions of privacy are required.

\subsection{Provable Private Release of Graph Statistics using Differential Privacy}
\label{subsec:provable graph query}

Provable privacy techniques provide mathematical guarantees about what an analyst can learn from the answers to a series of queries about a data set. These techniques differ from the traditional methods by precisely quantifying and controlling the amount of information that can be leaked using tunable parameters. Among all provable privacy techniques, perhaps the most well studied and widely recognized privacy definition is Differential Privacy (\Cref{def:dp}). Since the paper of \citet{dwork2006calibrating} that formalizes the concept of noise addition according to the sensitivity of a query function, there have been enormous studies on DP and its applications in the context of tabular data. It was inevitable that DP would be employed to manage privacy concerns in graph data, starting from an early work that introduces smooth sensitivity \cite{nissim2007smooth}. The key to graph data is the representation and storage of records in graphs, in which nodes represent data set entities and edges represent relationships between entities. This also gives rise to two distinct concepts of graph differential privacy -- edge differential privacy and node differential privacy -- that will be discussed in detail in the following sections. 

The fundamental difference between the two versions of graph differential privacy is how a pair of neighboring graphs is defined. 
In the standard form of DP (e.g., in \cite{dwork2006calibrating} and \cite{dwork2014algorithmic}), two data sets $x$ and $y$ are neighbors if they differ by at most one record. \citet{hay2009accurate} formalizes the concepts of edge and node differential privacy for graphs by generalizing the definition of neighboring data sets using the \textit{symmetric difference} between two sets. The symmetric difference $x \oplus y$ between two sets $x$ and $y$ (not necessarily the same size) is the set of elements in either $x$ or $y$, but not in both, i.e., $x \oplus y = (x \cup y) \setminus (x \cap y)$.   
With this, neighboring graphs in the context of edge and node differential privacy are defined as follows:

\begin{definition}
\label{definition: edge neighboring graph}
Given a graph $G=(V, E)$, a graph $G'=(V',E')$ is an \textbf{edge neighboring graph} of $G$ if it differs from $G$ by exactly one edge, i.e., $|V \oplus V'|+|E \oplus E'|=1$.
\end{definition}

\begin{example} 
The graphs in \Cref{subfig:a graph} and \Cref{subfig:edge nbr graph} are edge neighboring graphs because $|V \oplus V'| + |E \oplus E'| = 0 + 1 = 1$, i.e., they differ by exactly one edge $12$. 
\end{example}

\begin{definition}
\label{definition: node neighboring graph}
Given a graph $G=(V, E)$, a graph $G'=(V',E')$ is a \textbf{node neighboring graph} of $G$ if it differs from $G$ by exactly one node and the edges incident to the node, i.e., $|V \oplus V'|=1$ and $E \oplus E' = \{uv \mid u = V \oplus V' \text{ or } v = V \oplus V'\}$.
\end{definition}
\begin{example} 
The graphs in \Cref{subfig:a graph} and \Cref{subfig:node nbr graph} are node neighboring graphs because $|V \oplus V''| = 1$ and $E \oplus E'' = \{32, 34, 35\}$, i.e., they differ by exactly one node $3$ and the edge differences are the edges incident to $3$ in $G$.
\end{example}
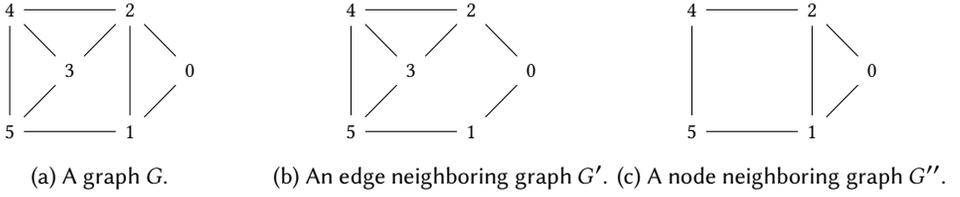
\begin{figure}[t!]
\centering
\begin{subfigure}[t]{0.32\textwidth}
\centering
\begin{tikzpicture}[scale=0.8]
\begin{scope}[>={Stealth[black]},every edge/.style={draw=black}]
    \node (0) at (2,0) {0};
    \node (1) at (1,-1) {1};
    \node (2) at (1,1) {2};
    \node (3) at (0,0) {3};
    \node (4) at (-1,1) {4};
    \node (5) at (-1,-1) {5};
    \path [-] (0) edge (1);
    \path [-] (0) edge (2);
    \path [-] (1) edge (5);
    \path [-] (1) edge (2);
    \path [-] (2) edge (3);
    \path [-] (2) edge (4);
    \path [-] (3) edge (4);
    \path [-] (3) edge (5);
    \path [-] (4) edge (5);
\end{scope}
\end{tikzpicture}
\caption{A graph $G$.}
\label{subfig:a graph}
\end{subfigure}
\begin{subfigure}[t]{0.32\textwidth}
\centering
\begin{tikzpicture}[scale=0.8]
\begin{scope}[>={Stealth[black]},every edge/.style={draw=black}]
    \node (0) at (2,0) {0};
    \node (1) at (1,-1) {1};
    \node (2) at (1,1) {2};
    \node (3) at (0,0) {3};
    \node (4) at (-1,1) {4};
    \node (5) at (-1,-1) {5};
    \path [-] (0) edge (1);
    \path [-] (0) edge (2);
    \path [-] (1) edge (5);
    \path [-] (2) edge (3);
    \path [-] (2) edge (4);
    \path [-] (3) edge (4);
    \path [-] (3) edge (5);
    \path [-] (4) edge (5);
\end{scope}
\end{tikzpicture}
\caption{An edge neighboring graph $G'$.}
\label{subfig:edge nbr graph}
\end{subfigure}
\begin{subfigure}[t]{0.32\textwidth}
\centering
\begin{tikzpicture}[scale=0.8]
\begin{scope}[>={Stealth[black]},every edge/.style={draw=black}]
    \node (0) at (2,0) {0};
    \node (1) at (1,-1) {1};
    \node (2) at (1,1) {2};
    \node (4) at (-1,1) {4};
    \node (5) at (-1,-1) {5};
    \path [-] (0) edge (1);
    \path [-] (0) edge (2);
    \path [-] (1) edge (5);
    \path [-] (1) edge (2);
    \path [-] (2) edge (4);
    \path [-] (4) edge (5);
\end{scope}
\end{tikzpicture}
\caption{A node neighboring graph $G''$.}
\label{subfig:node nbr graph}
\end{subfigure}
\caption{Examples of an edge neighboring graph and a node neighboring graph.}
\label{fig:neighboring graphs}
\end{figure}

\noindent By generalizing \Cref{definition: edge neighboring graph} to multiple edge differences, Hay et al.~\cite{hay2009accurate} states the following. 

\begin{definition}
Given a graph $G=(V, E)$, a graph $G'=(V',E')$ is a \textbf{$k$-edge neighboring graph} of $G$ if it differs from $G$ by at most $k$ edges. That is, $|V \oplus V'|+|E \oplus E'| \le k$.
\end{definition}
The connection between $k$-edge differential privacy and node differential privacy depends on $k$ and node degrees in a graph. If $k$ is larger than the maximum degree in the graph, then $k$-edge differential privacy is stronger than node differential privacy. Otherwise, it may simultaneously protect multiple relationships for one node or several nodes. 
With the three versions of neighboring graphs given above, it is straightforward to formalize  edge and node differential privacy based on \Cref{def:dp for graph}.

Generally speaking, it is more difficult to satisfy differential privacy on graph data than on tabular data, because graph queries typically have higher sensitivities than statistical queries.
The two versions of graph differential privacy address different privacy concerns. 
Edge differential privacy protects the relationship between two entities while node differential privacy protects the existence of an entity and its relationships with others.
Frequently, edge differential privacy is easier to achieve than node differential privacy.

\begin{example}
Consider the graph $G$ which is depicted in \Cref{subfig:a graph}. It has the degree sequence $(4,3,3,3,3,2)$. It has an edge neighboring graph $G^{\prime}$ and node neighboring graph $G^{\prime\prime}$ which are depicted in  \Cref{subfig:edge nbr graph} and \Cref{subfig:node nbr graph} respectively. Notice that $G^{\prime}$ has the degree sequence $(3,3,3,3,2,2)$ and $G^{\prime\prime}$ has the degree sequence $(3,3,2,2,2,0)$. So the sensitivities of the degree sequence function for the graph $G$ under the edge and node differential privacy frameworks are $2$ and $6$, respectively.  
\end{example}

\subsubsection{Edge Differential Privacy} 
\label{section:edge privacy}

As discussed above, edge differential privacy can protect relationships between entities in a network from a malicious analyst who can query that network. These queries usually have high sensitivities due to their unique nature. In spite of that, there are many examples of mechanisms that preserve edge differential privacy and that can be used to release a wide variety of graph statistics. These include the protections of 
edge weight \cite{costea2013qualitative,li2017differential} which may reflect communication frequency,  
vertex clustering coefficient \cite{wang2013learning} for analyzing a vertex's connectivity to its neighbors, 
eigenvalues and eigenvectors \cite{wang2013differential,ahmed2019publishing} for analyzing characteristics of network adjacency matrices,
and egocentric betweenness centrality \cite{roohi2019differentially} for analyzing the importance of a vertex linking two parts of a network, 
community detection \cite{nguyen2016detecting}. There are also techniques that preserve variations of edge differential privacy, such as
graph clustering under a weaker version of $k$-edge differential privacy \cite{mulle2015privacy} and 
popularity graphs under outlink privacy \cite{task2012guide}.

In the rest of this subsection, we focus on reviewing published works for subgraph counting, degree sequence/distribution, and cut queries. We focus on these classes of queries because they have been relatively well studied due to their usefulness for other related studies. 
The main building blocks of edge differentially private mechanisms are the Laplace and the exponential mechanisms. The former is used in combination with different types of sensitivities to calculate the right amount of noise that needs to be added. The latter is often used for parameter selection or non-numerical outputs.

\paragraph{Subgraph Counting}
Subgraph counting queries count the number of times a certain subgraph appears in a given graph. Common subgraphs include triangles and stars, and their generalization to $k$-triangles (i.e., a subgraph consists of $k$ triangles, all of which share a common edge) and $k$-stars (i.e., a subgraph with $k+1$ nodes in which a central node has degree $k$ and the other $k$ nodes have degree 1). 
The work in \cite{nissim2007smooth} is one of the earliest on making subgraph-counting queries satisfy edge differential privacy. It introduces the notion of smooth sensitivity for a given graph, which can be used in place of global sensitivity (\Cref{equation:global sensitivity}) in order to reduce the amount of noise required to ensure that the queries preserve DP. 

\begin{definition}
Given a data set $D_x \in D^n$, for a real-valued function $f: D^n \rightarrow \mathbb{R}^1$, the \textbf{local sensitivity} of $f$ at $D_x$ is 
\begin{equation*}
    LS_f(D_x) = \max_{D_y \in D^n \,:\, d(D_x,D_y)=1} ||f(D_x) - f(D_y)||_1.
\end{equation*}
\end{definition}

\begin{figure}[t!]
\centering
\begin{subfigure}[t]{0.4\textwidth}
\centering
\begin{tikzpicture}[scale=0.8]
\begin{scope}[>={Stealth[black]}]
    \node (0) at (2,0) {0};
    \node (1) at (1,-1) {1};
    \node (2) at (1,1) {2};
    \node (3) at (0,0) {3};
    \node (4) at (-1,1) {4};
    \node (5) at (-1,-1) {5};
    \node (6) at (-3,1) {6};
    \node (7) at (-3,-1) {7};
    \path [-] (0) edge (1);
    \path [-] (0) edge (2);
    \path [-] (1) edge (2);
    \path [-] (1) edge (5);
    \path [-] (2) edge (3);
    \path [-] (2) edge (4);
    \path [-] (3) edge (4);
    \path [-] (3) edge (5);
    \path [-, blue] (4) edge (5);
    \path [-] (4) edge (6);
    \path [-] (4) edge (7);
    \path [-] (5) edge (6);
    \path [-] (5) edge (7);
    \path [-] (6) edge (7);
\end{scope}
\end{tikzpicture}
\caption{A graph $G$ with 7 triangles. Deleting the edge $45$ results in an edge neighboring graph with 4 triangles. So the local sensitivity at $G$ is 3.}
\label{subfig:triangle count on g}
\end{subfigure}\quad
\begin{subfigure}[t]{0.4\textwidth}
\centering
\begin{tikzpicture}[scale=0.8]
\begin{scope}[>={Stealth[black]}]
    \node (0) at (2,0) {0};
    \node (1) at (1,-1) {1};
    \node (2) at (1,1) {2};
    \node (3) at (0,0) {3};
    \node (4) at (-1,1) {4};
    \node (5) at (-1,-1) {5};
    \node (6) at (-3,1) {6};
    \node (7) at (-3,-1) {7};
    
    \path [-] (0) edge (1);
    \path [-] (0) edge (2);
    \path [-] (1) edge (2);
    \path [-] (1) edge (5);
    \path [-] (2) edge (3);
    \path [-] (2) edge (4);
    \path [-] (3) edge (4);
    \path [-] (3) edge (5);
    \path [-, blue] (4) edge (5);
    \path [-] (4) edge (6);
    \path [-] (4) edge (7);
    \path [-] (5) edge (6);
    \path [-] (5) edge (7);
    \path [-] (6) edge (7);
    
    \path [bend right, green] (5) edge (2);
    \path [bend left,green] (4) edge (1);
\end{scope}
\end{tikzpicture}
\caption{A graph $G'$ s.t. $d(G,G')=2$. $G'$ has 12 triangles. Deleting the edge $45$ results in an edge neighboring graph with 7 triangles. So the local sensitivity at $G'$ is 5.}
\label{subfig:triangle count on g nbr}
\end{subfigure}
\caption{Triangle counting query and its local sensitivity.}
\label{fig:triangle count}
\end{figure}
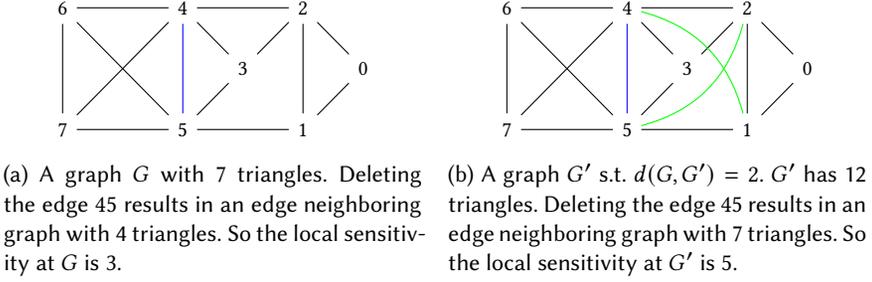

In the worst case, the local sensitivity of a function is the same as the global sensitivity but it can be much smaller for a given data set $D_x$.  
For triangle-counting queries, the global sensitivity is $|V| - 2$, but the local sensitivity is $\max_{i,j \in [n]} a_{ij}$, where $a_{ij}$ is the number of common neighbors between adjacent nodes $i$ and $j$.  
For example, the graph $G$ in \Cref{subfig:triangle count on g} has 7 triangles. The local and global sensitivities of the triangle counting queries at $G$ are 3 and 6, respectively. 

Unfortunately, using local sensitivity directly can reveal sensitive information about the underlying graph. For example, the local sensitivity of a triangle-counting query gives the maximum number of common neighbors between two vertices in the given graph. A more suitable candidate is the smallest smooth upper bound of the local sensitivity, namely the \textit{$\beta$-smooth sensitivity} $SS_{f,\beta}$. For $\beta > 0$ we have
\begin{equation}
\label{equation:smooth sensitivity}
    SS_{f,\beta}(D_x) = \max_{D_y \in D^n} \left(LS_f(D_y) \cdot \exp^{-\beta d(D_x,D_y)} \right).
\end{equation}
Intuitively, the smooth sensitivity of a function $f$ of a data set $D_x$ comes from a data set $D_y$ that is close to $D_x$ and also has a large local sensitivity. 

\begin{example}
To calculate the smooth sensitivity of the triangle-counting query for the graph $G$ in \Cref{subfig:triangle count on g}, we go through all $k$-edge-neighbors of $G$, calculate their local sensitivities, then pick the one that satisfies \Cref{equation:smooth sensitivity}. In this case, the smooth sensitivity $SS_{f,0.1} \approx 4.1$ happens at a neighboring graph $G'$ as shown in \Cref{subfig:triangle count on g nbr}, where $d(G,G')=2$ and $LS_f(G') = 5$.
\end{example}

The advantage of using smooth sensitivity is demonstrated in \cite{nissim2007smooth} through a few examples, including privately releasing the cost of the minimum spanning tree and the number of triangles in a graph. Since then, smooth sensitivity has been used widely in subsequent works to improve the utility of differentially private query responses, including many works in graph differential privacy. 

Two mechanisms based on local sensitivity are proposed in \cite{karwa2011private} to release $k$-star counts and k-triangle counts. 
The first mechanism is a direct extension from \cite{nissim2007smooth} to $k$-star counting queries. It is achieved by an efficient algorithm to compute the smooth sensitivity in time  $O(n \log n + m)$, where $n$ and $m$ are the numbers of nodes and edges in a graph respectively. 
The second mechanism relies on a bound on the smooth sensitivity rather than an efficient algorithm to compute its exact value. For this mechanism, the local sensitivity of the $k$-triangle counting queries for $k \ge 2$ is masked by its local sensitivity, which gives the second order local sensitivity $LS_f'$ with a simple upper bound 
\begin{equation}
\label{equation:kawar ktriangle 2nd ls upper bound}
    LS_f'(G) \le 3 \binom{a_{max}}{k - 1} + a_{max} \binom{a_{max}}{k-2},  
\end{equation}
where $a_{max}$ is the maximum number of common neighbors between a pair of vertices in $G$. 
The mechanism releases the true k-triangle count with Laplace noise proportional to the above upper bound and runs in time $O(md)$, where $m$ is the number of edges and $d$ is the maximum degree. 

\begin{example}
Let $k=2$ and observe that the number of $2$-triangles in \Cref{subfig:triangle count on g} is 9. A closed form equation in \cite{karwa2011private} shows that the local sensitivity at $G$ is 8, which happens when deleting the edge $45$. Since $a_{max} = a_{45}=3$, the upper bound $LS'(G) \le 12$ by \Cref{equation:kawar ktriangle 2nd ls upper bound}. So the mechanism output is $8$ plus Laplace noise (proportional to 12) plus an additional term. 
\end{example}

As stated in \cite{karwa2011private}, the released answers for 2-stars and 3-stars are useful in moderately dense graphs, and for triangles and 2-triangles the answers are useful in dense graphs but not so for sparse graphs.
An application of private k-triangle counting appears in \cite{lu2014exponential}, where the authors use it in conjunction with private alternating \textit{k-star} counting and private alternating \textit{k-twopath} counting to estimate the parameters for exponential random graph models.   

In contrast to the Laplace-based approaches, a different method for subgraph counting based on the exponential mechanism is proposed in \cite{zhang2015private}. In this work, the authors use a ladder function as the utility function. 
Under certain conditions of the ladder function, the proposed mechanism is differentially private. 
An optimal choice of a ladder function is the generalized \textit{local sensitivity at distance $t$} (i.e., the maximum change of a function's output between a given data set $D^{(1)}$ and a neighboring data set $D^{(2)}$ such that $d(D^{(1)},D^{(2)}) \le t$).  
This function is only efficiently computable for triangle and k-star counting queries. For $k$-clique and  $k$-triangle counting queries, a more efficient choice is a convergent upper bound of the local sensitivity at distance $t$.
Empirical evaluations in \cite{zhang2015private} on real graph data show substantial improvements in the exponential-mechanism-based methods over the Laplace-mechanism-based methods with global sensitivity, smooth sensitivity \cite{nissim2007smooth}, second order local sensitivity \cite{karwa2011private} and a recursive strategy \cite{chen2013recursive}. 

To reduce the noise effect in the worst case, \citet{proserpio2014calibrating} scales down the influence of \emph{troublesome} records in weighted data sets using a platform called \textit{weighted Privacy Integrated Query} (wPINQ), which is built upon the \textit{PINQ} \cite{mcsherry2009privacy} platform that guarantees all acceptable queries satisfy differential privacy based on the Laplace and exponential mechanisms. 
In wPINQ, weights are scaled down differently for various built-in operators such as \textit{Select}, \textit{Join} and \textit{GroupBy}. 
An earlier work \cite{proserpio2012workflow} demonstrates how wPINQ can produce private degree distributions and joint degree distributions of weighted graphs for the purpose of generating private synthetic graphs. 
\citet{proserpio2014calibrating} then demonstrates wPINQ for triangle by degrees counts and square by degrees counts, and shows how these statistics can be combined with Markov Chain Monte Carlo (MCMC) to generate private synthetic graphs. 

\paragraph{Degree Sequence}

Another widely studied graph statistic is the degree sequence of a graph. Give a graph $G$, its \textit{degree sequence} $S(G)$ is a monotonic non-decreasing sequence of node degrees.
It can be used to compute the average or maximum degree of a graph. It can also be used to recover a graph, provided there is a consistent graph with the given degree sequence. 
A degree sequence can be transformed into a \textit{degree distribution}, which is a useful feature for graph classification, e.g., the degree distribution of a scale-free network follows a power law. 
When perturbing the existence of an edge in a given graph, its degree sequence's global sensitivity is two, which behaves much better than that of subgraph-counting query functions. 

\citet{hay2009accurate} proposes a constraint inference-based algorithm that can be used as a post-processing step to improve the quality of degree sequences released using private mechanisms.
Given a noisy degree sequence $\Tilde{S}(G)$ produced by a private mechanism, constraint inference finds a non-increasing ordered degree sequence $\Bar{S}(G)$ (in the same vertex order) based on \textit{isometric regression} such that the difference $||\Tilde{S}(G)-\Bar{S}(G)||_2$ is minimized. 
\begin{example}\label{ex:constraint inference} 
The graph in \Cref{subfig:triangle count on g} has the degree sequence $S = (2,3,3,3,3,4,5,5)$ a possible private degree sequence is $\Tilde{S}(G) = (3,2,5,3,-4,16,6,3)$, produced by the Laplace mechanism with noise sampled from $Lap(\frac{2}{0.5})$. The distance between these two 
is $||S(G) - \Tilde{S}(G)||_2 \approx 14$. 
After applying constraint inference to $\Tilde{S}(G)$, the post-processed degree sequence is 
$\overline{S}(G) = (2,2,2,2,2,8,8,8)$ with distance reduced to $||S(G) - \overline{S}(G)||_2 \approx 6$. \end{example}
Constraint inference is commonly adopted in subsequent works to improve query utility.  
The advantages of constraint inference are its computational efficiency (100 million nodes were processed in just a few seconds in \cite{hay2009accurate}), its applicability to a wide range of graph statistics, and the fact that its error increases only linearly with the number of unique degrees. 
An issue with constraint inference is that the post-processed degree sequence can be non-graphical, i.e., not consistent with any graph. In \Cref{ex:constraint inference}, the post-processed degree sequence $(2,2,2,2,2,8,8,8)$ is non-graphical. This could be a problem if the private degree sequence is used to carry out statistical inference or generate synthetic graphs. To overcome this, \cite{karwa2012differentially} presents an additional optimization step after constraint inference in the domain of all graphical degree sequences. 

A mechanism that produces both the private degree distribution of a graph and the joint degree distribution of a graph is the aforementioned \cite{proserpio2012workflow}, which is essentially based on the Laplace and exponential mechanisms.

\paragraph{Cut Query}

Sometimes one is interested in the number of interactions between two groups of entities in a network, for example the number of sales interactions between two companies or the number of collaborations between two groups of researchers from different organizations. In graph data, these can be measured by cut queries between two vertex subsets. 
More precisely, given a weighted graph $G=(V,E)$ and two non-empty subsets $V_S, V_T \subseteq V$, an s-t-cut query returns the total weight of the edges crossing $V_S$ and $V_T$. 

\citet{gupta2012iterative} employs an \textit{iterative database construction} (IDC) framework that generalizes the \textit{median mechanism} \cite{roth10} and \textit{multiplicative weights mechanism} \cite{hardt10} with tighter bounds. 
It iteratively compares the response $Q^t(D^t)$ from an approximated data set $D^t$ with a noisy response $Q^t(D)+Lap(\cdot)$ from the given data set $D$, and updates the current approximation to $D^{t+1}$ if the two responses are not close. 
However, it requires graphs that are sufficiently dense.

A different approach is taken in \cite{blocki2012johnson} to release answers of cut queries with less additive noise. They show that one can apply the 
Johnson-Lindenstrauss (JL) transform \cite{johnson1984extensions} 
to an updated edge matrix of a weighted graph to obtain a \textit{sanitized} graph Laplacian matrix. This sanitized Laplacian matrix can then be used to approximate s-t-cut queries while preserving $(\epsilon,\delta)$-edge differential privacy. 
Before applying the JL transform, each weight $w_{uv}$ in the edge matrix $M_E$ is updated by $w_{uv} = \frac{w}{n} + (1-\frac{w}{n}) w_{uv}$ where $n$ is the number of vertices and $w$ is calculated from pre-determined parameters. 
Once the update is complete, the JL transform is applied by the step 
$\overline{L} = \frac{1}{r}M_E^T\cdot M^T\cdot M\cdot M_E$, where entries of the matrix $M$ are sampled \textit{i.i.d.} from $N(0,1)$.
This method only adds (with high probability) constant noise (w.r.t. graph size) to cut query answers, hence provides superior results compared to \cite{gupta2012iterative} for small cuts $\le O(n)$.  
In fact, the JL transform-based strategy can serve as a more general approach to publishing randomized graphs that satisfy edge differential privacy for any downstream graph queries. 

The above JL transform-based method is revised for sparse graphs by \cite{Upadhyay2013}. 
The authors observe that the weight update in \cite{blocki2012johnson} is for all possible pairs of vertices, which is the same as overlaying a complete graph on $G$. 
This is done to ensure the graph corresponding to the updated edge weights is well connected prior to applying the JL transform.  
To maintain the sparsity of the given graph $G$, a $d$-regular expander graph $E$ is used when performing weight update in $G$ by $L_H = \frac{w}{d} L_E + (1-\frac{w}{d}) L_G$. (An expander graph is a sparse graph such that every small vertex subset is well connected to the rest of the graph.) This is followed by a step of Gaussian noise addition to ensure differential privacy. 
More generally, combining this sanitization process with a graph sparsification technique preserves graph differential privacy for cut queries for arbitrary graphs.


\subsubsection{Node Differential Privacy}
\label{sec:nodepriv}

The distinction between node and edge differential privacy originates from how a neighboring graphs are defined. In the context of node differential privacy, a pair of neighboring graphs differ by exactly one node and the edges incident to the node (\Cref{definition: node neighboring graph}). 
The advantage of this is that it gives a higher level of protection about an entity's privacy, including its existence in the data set and its relations to others.
The disadvantage is it tends to give rise to high query sensitivity.
Hence, it is more difficult to achieve node differential privacy with useful query responses. 
Today, there are two main types of solutions that deal with such high query sensitivity. The first is based on a top-down projection to lower degree graphs. The second is based on a bottom-up generalization using Lipschitz extensions. 

The top-down methods project the given graph to a graph with a bounded maximum degree, then answer queries using the projected graph with noise that is proportional to the sensitivities of the query function and the projection map.
A projection could be a na{\"i}ve removal of all high degree nodes or a random edge removal that reduces high node degrees.
These projections are generic and easy to implement, but suffer from high sensitivity and potential information loss.

The bottom-up methods first answer queries on bounded degree graphs, then extend the answers to arbitrary graphs using Lipschitz extensions. The extended answers can then be released with additive noise according to existing differential privacy mechanisms, such as the Laplace mechanism. 
The main drawback is designing an appropriate Lipschitz extension for each query function, which is a non-trivial task in general. 
Other than that, Lipschitz extension-based methods are usually preferred because of their generalizations to arbitrary graphs to avoid information loss. 

For the same reason as explained in \Cref{section:edge privacy}, we only review published works for releasing degree sequence/distribution and subgraph counting queries. Due to the difficulty in obtaining high utility private mechanisms, there are fewer works concerning node differential privacy than there are concerning edge differential privacy. That said, some recent works have studied private estimation of generative graph model parameters \cite{borgs2015private,borgs2018revealing,sealfon2019efficiently}, such as the  edge probability in the Erd\H{o}s-R\'{e}nyi model.

Sometimes, instead of a static graph, there may be a sequence of dynamic graphs. In this case, one may want to release some common graph statistics such as degree sequence or subgraph counts for the entire graph sequence. A study in \cite{song2018differentially} presents a mechanism to publish \textit{difference sequence}, which is the difference between the graph statistic for two graphs adjacent in time.

Before proceeding, we give the definitions of Lipschitz function and Lipschitz extension. Intuitively, a function is Lipschitz if the \emph{difference} between its images is bounded by a constant times the \emph{difference} between the corresponding preimages. Formally, it is defined as: 

\begin{definition}[Lipschitz constant]
Given two metric spaces $(X, d_X)$ and $(Y, d_Y)$, a function $f:X \rightarrow Y$ is called \textbf{$c$-Lipschitz} (or has a \textbf{Lipschitz constant} $c$) if there exists a real constant $c \ge 0$ such that for all $x_1, x_2 \in X$, 
    $d_Y(f(x_1),f(x_2)) \le c \cdot d_X(x_1, x_2).$
\end{definition}

In the context of graph differential privacy, the graph space with node or edge edit distance as the metric is mapped by the function $f$ to a real valued space with, for example, $L_1$-norm as the metric. Choices of $f$ include degree sequence, subgraph counting, node centrality score, etc. 
The global sensitivity of a function is the smallest Lipschitz constant that upper bounds the maximum changes in the codomain. 

A $c$-Lipschitz function may be extended to another Lipschitz function that takes on a larger domain with the same codomain. Such an extension is called a Lipschitz extension and is essential for extending a node differentially private query-answering mechanism from a restricted graph domain (e.g., bounded degree graphs) to the general graph domain.  

\begin{definition}[Lipschitz extension]
Given two metric spaces $(X, d_X)$ and $(Y, d_Y)$ and a $c$-Lipschitz function $f':X' \rightarrow Y$ with the domain $X' \subseteq X$, a function $f:X \rightarrow Y$ is a \textbf{Lipschitz extension} of $f'$ from $X'$ to $X$ with stretch $s \ge 1$ if 
\begin{enumerate}
    \item the two functions $f(x)=f'(x)$ are identical for all $x \in X'$ and 
    \item the extended function $f$ is $sc$-Lipschitz. 
\end{enumerate}
\end{definition}

The desire for high-utility node differentially private mechanisms motivates the search for efficiently computable Lipschitz extensions with low Lipschitz constants and stretches from restricted graphs to arbitrary graphs. The following subsections review some results in this area, with a focus on degree-sequence and subgraph-counting queries, using both the top-down projection and the bottom-up Lipschitz-extension approaches.   

\paragraph{Degree Sequence}

\citet{kasiviswanathan2013analyzing} 
describes a mechanism that privately releases the degree distribution of a graph using a simple projection function $f_T:G \rightarrow G_\theta$ that discards all nodes in $G$ whose degrees are higher than a  threshold $\theta$. 
They showed that if $U_s(G)$ is a smooth upper bound on the local sensitivity of $f_T$ and $\Delta_{\theta}f$ is the global sensitivity of a query $f$ on graphs with bounded maximum degree $\theta$, then the function composition $f \circ f_T$ has a smooth upper bound $U_s(G) \cdot \Delta_{\theta}f$. 
\citet{kasiviswanathan2013analyzing} gave explicit formulas for computing $\Delta_{\theta}f$ and the local and smooth sensitivities of the truncation function $f_T$.
Furthermore, it was proved that randomizing the truncation cutoff in a range close to the given threshold $\theta$ is likely to reduce its smooth sensitivity.
The presented mechanism was proved to satisfy node differential privacy with Cauchy noise. 
It runs in time $O(|E|)$ and produces private degree distributions with $L_1$ error $O\left(\frac{\Bar{d}^{\alpha} \ln n \ln \theta}{\epsilon^2 \theta^{\alpha-2}} + \frac{\theta^3 \ln \theta}{n \epsilon^2}\right)$ for graphs with $n$ nodes and average degree $\Bar{d}$, provided the graphs satisfy certain constraints and $\alpha$-decay, which is a mild assumption on the tail of the graph's degree distribution. The authors proved that if $\alpha>2$ and $\Bar{d}$ is polylogarithmic in $n$, then this error goes to 0 as $n$ increases. 

The na{\"i}ve truncation of \cite{kasiviswanathan2013analyzing} suffers from high local sensitivity due to the deletion of a large number of edges, especially in dense graphs. To address this, a projection based on edge addition is proposed in \cite{day2016publishing} for releasing private degree histograms. The authors prove that by adding edges to the empty graph in a stable edge ordering, the final graph not only has more edges than the one obtained through na{\"i}ve projection, but is maximal in terms of edge addition. 
Given two (node) neighboring graphs, an edge ordering of a graph is \textit{stable} if for any pair of edges that appears in both graphs, their relative ordering stays the same in the edge ordering of both graphs. 
The mechanism for releasing degree histograms employs the exponential mechanism to choose the optimal degree threshold $\theta$ and bin aggregation $\Omega$ for reducing sensitivity, then adds $Lap(\frac{2\theta+1}{\epsilon_2})$ noise to the aggregated bins. 
This mechanism satisfies $(\epsilon_1+\epsilon_2)$-node differential privacy, where $\epsilon_1$ and $\epsilon_2$ are the privacy budgets for the exponential and Laplace mechanisms respectively. It runs in time $O(\Theta \cdot |E|)$, where $\Theta$ is an integer upper bound of $\theta$. An extension that releases cumulative histograms require only $Lap(\frac{\theta+1}{\epsilon})$ noise and has the additional benefit that the result can be post-processed by the constraint-inference algorithm \cite{hay2009accurate}. The extension relies on the exponential mechanism to select an optimal $\theta$ so it has the same time complexity and privacy bound. 

A key method based on the Lipschitz extension for releasing private degree sequence is \cite{raskhodnikova2016lipschitz}. Since the global sensitivity for bounded degree graphs $G_\theta$ satisfies $\Delta_{S(G_{\theta})} \le 2\theta$ , it implies that $S(G_{\theta})$ has a Lipschitz constant $2\theta$. By constructing a flow graph $FG(G)$ from the given graph $G$ and a degree threshold $\theta$, \citet{raskhodnikova2016lipschitz} present a Lipschitz extension of $S(G_{\theta})$ from the set of bounded-degree graphs to arbitrary graphs with low stretch via a strongly convex optimization problem. 
More specifically, the flow graph $FG(G)$ has a source node $s$, a sink node $t$ and two sets of nodes $V_l$ and $V_r$ that are exact copies of the nodes in $G$. The source $s$ is connected to all of $V_l$ via directed edges with capacity $\theta$, and similarly for $V_r$ and $t$. Each node $x \in V_l$  is connected to a node $y \in V_r$ via a directed edge with capacity 1 if there is an edge $xy$ in $G$. 
Given the flow graph, \cite{raskhodnikova2016lipschitz} solves for an optimal flow $f$ that minimizes the objective function $\Phi(f)=||(f_{s\cdot},f_{\cdot t})-\Vec{\theta}||_2^2$, where $f_{s\cdot}$ and $f_{\cdot t}$ are the vectors of flows leaving $s$ and entering $t$ respectively and $\Vec{\theta} = (\theta,\ldots,\theta)$ has length $2n$. 
The authors prove that the sorted $f_{s\cdot}$ is an efficiently computable Lipschitz extension of the degree-sequence function with a stretch of $1.5$. 

The same flow graph construction is used in \cite{kasiviswanathan2013analyzing} for subgraph counting queries. 
The result of \cite{raskhodnikova2016lipschitz} is consistent with \cite{kasiviswanathan2013analyzing}'s Lipschitz extension on edge count, but \cite{raskhodnikova2016lipschitz} minimizes the above objective function $\Phi(f)$ rather than maximize the network flow, because the maximum flow may not be unique and the two formulations have different sensitivities. 

With some additional work that replaces the scoring function in the standard exponential mechanism, \cite{raskhodnikova2016lipschitz} also proves that using the adjusted exponential mechanism to select the degree threshold leads to a Lipschitz extension with low sensitivity, and hence better utility in the private outputs.  
Combining the Lipschitz extension and the adjusted exponential mechanism, the authors are also able to release private degree distributions with improvement on the error bound of  \cite{kasiviswanathan2013analyzing}. 

\begin{figure}[t!]
\centering
\begin{tikzpicture}[scale=0.65]
	\begin{scope}[>={Stealth[black]},every edge/.style={draw=black}]
	\node (s) at (-5,0) {s};
	\node (t) at (5,0) {t};
	
	\node (l0) at (-2,2.5) {0};
	\node (l1) at (-2,1.5) {1};
	\node (l2) at (-2,0.5) {2};
	\node (l3) at (-2,-0.5) {3};
	\node (l4) at (-2,-1.5) {4};
	\node (l5) at (-2,-2.5) {5};
	
	\node (r0) at (2,2.5) {0};
	\node (r1) at (2,1.5) {1};
	\node (r2) at (2,0.5) {2};
	\node (r3) at (2,-0.5) {3};
	\node (r4) at (2,-1.5) {4};
	\node (r5) at (2,-2.5) {5};
	
	\node (label) at (0,2.7) {$1/1$};
	
	\path [->] (s) edge node[fill=white,above,sloped,right] {$2/4$}(l0);
	\path [->] (s) edge node[fill=white,above,sloped,right] {$3/4$}(l1);
	\path [->] (s) edge node[fill=white,above,sloped,right] {$4/4$}(l2);
	\path [->] (s) edge node[fill=white,above,sloped,right] {$3/4$}(l3);
	\path [->] (s) edge node[fill=white,above,sloped,right] {$3/4$}(l4);
	\path [->] (s) edge node[fill=white,above,sloped,right] {$3/4$}(l5);
	
	\path [->] (r0) edge node[fill=white,above,sloped,left] {$2/4$}(t);
	\path [->] (r1) edge node[fill=white,above,sloped,left] {$3/4$}(t);
	\path [->] (r2) edge node[fill=white,above,sloped,left] {$4/4$}(t);
	\path [->] (r3) edge node[fill=white,above,sloped,left] {$3/4$}(t);
	\path [->] (r4) edge node[fill=white,above,sloped,left] {$3/4$}(t);
	\path [->] (r5) edge node[fill=white,above,sloped,left] {$3/4$}(t);
	
	\path [->] (l0) edge (r1);
	\path [->] (l0) edge (r2);
	\path [->] (l1) edge (r0);
	\path [->] (l1) edge (r2);
	\path [->] (l1) edge (r5);
	\path [->] (l2) edge (r0);
	\path [->] (l2) edge (r1);
	\path [->] (l2) edge (r3);
	\path [->] (l2) edge (r4);
	\path [->] (l3) edge (r2);
	\path [->] (l3) edge (r4);
	\path [->] (l3) edge (r5);
	\path [->] (l4) edge (r2);
	\path [->] (l4) edge (r3);
	\path [->] (l4) edge (r5);
	\path [->] (l5) edge (r1);
	\path [->] (l5) edge (r3);
	\path [->] (l5) edge (r4);
	
	\end{scope}
	\end{tikzpicture}
\caption{The flow graph constructed from the graph in \Cref{subfig:a graph} with degree bound $\theta=4$. The maximum flow and edge capacities are shown on the edges. The arbitrary choice of $\theta$ can affect the final output accuracy.}
\label{fig:flow graph}
\end{figure}
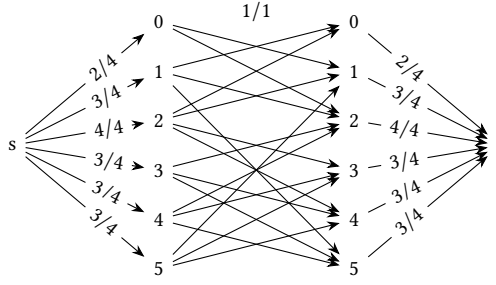

\begin{example}
Consider the graph $G$ shown in \Cref{subfig:a graph}.
Its flow graph is shown in \Cref{fig:flow graph}. 
The sorted out-flows is $f_{s\cdot}=(4,3,3,3,3,2)$, which is exactly the degree sequence of the original graph. The threshold $\theta$ is chosen to be $4$ on purpose, which matches the maximum degree in the original graph, so no information is lost during the problem transformation. 
\end{example}

\paragraph{Subgraph Counting}

The concept of restricted sensitivity is introduced in~\cite{blocki2013differentially}
to provide a top-down projection-based method for releasing private subgraph counts
This projection is the opposite of the edge-addition projection of \cite{day2016publishing} in the sense that it removes edges by following a canonical edge ordering. An edge $e$ is removed from the graph if it is incident to a vertex whose degree is higher than a threshold $\theta$ and $e$ is not the first $\theta$ edges in the ordering. 
The composition of the projection with a subgraph-counting query $f$ results in a function $f_{\mathcal{H}}$ that has global sensitivity (in the context of edge DP) and smooth sensitivity (in the context of node DP) proportional to the restricted sensitivity $RS_f(\mathcal{H}_{\theta})$ and $RS_f(\mathcal{H}_{2\theta})$ respectively, where $\mathcal{H}_{\theta}$ is the set of graphs with bounded degree $\theta$. The paper theoretically shows the advantage of restricted sensitivity in local-profile and subgraph-counting queries when adding Laplace noise, and gives explicit upper bounds for both query classes.

A key early work~\cite{kasiviswanathan2013analyzing} studies Lipschitz extension for private subgraph counting.
For edge-counting queries, flow graphs are constructed as described for \cite{raskhodnikova2016lipschitz} above (see also \Cref{fig:flow graph}). The maximum flow satisfies $f_{max}(G_{\theta})=2|E(G_{\theta})|$ and  $f_{max}(G) \le 2|E(G)|$ for bounded-degree graphs and arbitrary graphs, respectively. So $f_{max}$ is an efficiently computable Lipschitz extension of the edge-counting queries with global sensitivity $\Delta_{f_{max}} \le 2\theta$. 
To achieve $\epsilon$-node DP, one could then alternate between $|E|+Lap(\frac{2n}{\epsilon})$ and $\frac{f_{max}}{2}+Lap(\frac{2\theta}{\epsilon})$, depending on how close the former is to the truth. 
The same technique can be generalized to concave functions, where a function $h$ is concave if its increments $h(i+1)-h(i)$ are non-increasing as $i$ goes from $0$ to $n-2$. 

For small subgraph counting queries such as triangle counts, \cite{kasiviswanathan2013analyzing} proposes a linear programming formulation that maximizes $\sum x_C$ over all subgraphs of $G$ with some constraints, where $x_C=1$ if the subgraph of $G$ matches the given subgraph of interest and $x_C=0$ otherwise. The maximum value $v_{LP}(G)$ of this linear program satisfies the requirements of being a Lipschitz extension of subgraph counting queries $f_H(G)$ and has sensitivity bounded by $\Delta_{v_{LP}(G)} \le 6\theta^2$ for subgraphs with three nodes. Similar to private edge counts, small subgraph counts can also be released with $\epsilon$-node DP by alternating between $f_H(G)+Lap(\frac{6n^2}{\epsilon})$ and $v_{LP}(G)+Lap(\frac{6\theta^2}{\epsilon})$, depending on how close the former is to the truth. 

Another related work~\cite{ding2018privacy} applies a stable edge ordering scheme~\cite{day2016publishing} to a sequence of node pairs in order to obtain a projected graph, in which each node appears in a bounded number of triangles. The proposed private mechanisms can be used for releasing triangle count distributions, cumulative triangle count distributions and local clustering coefficient distributions. 

\subsubsection{Edge Weight Differential Privacy}
The previous two subsections focused on protecting the graph structures which may not be known to the public.  
When a graph structure is public and a system protects the edge weights or related statistics of the graph, such as distances between vertices, the aforementioned edge and node differential privacy frameworks may not be appropriate. Instead, a more suitable setting is to define differential privacy in the context of neighboring weight functions for the given graph. This is formally introduced in \cite{sealfon2016shortest}. A use case of this is a navigation system that has access to a public map and road user-traffic data and is required to keep user data private. Another use case is the World Wide Web as mentioned in \cite{pinot2018}. Two related but different works for protecting edge weight are \cite{costea2013qualitative} and \cite{li2017differential}. The former considers edge weights as counts from the data set, so the Laplace mechanism is used to protect counting queries.
The latter focuses on neighboring graphs that differ on at most one edge weight.

Given a graph and a weight function on the edges, the private edge weight differential privacy model in  \cite{sealfon2016shortest} is based on the following definition of neighboring weight functions.

\begin{definition}
\label{definition:neighboring weight function}
	Given a graph $G=(V,E)$, two weight functions $\omega, \omega': E \rightarrow \mathbb{R}^+$ are \textbf{neighboring} if the total weight difference is at most 1, i.e.,   
	\begin{equation*}
		||\omega - \omega'||_1 := \sum_{e \in E} |\omega(e) - \omega'(e)| \le 1.
	\end{equation*}
\end{definition}

Under this new framework, the weight change in a pair of edge neighboring graphs is at most 1, so \cite{sealfon2016shortest} uses the Laplace mechanism to release the distance between a pair of nodes and approximate distances between all pairs of nodes. 
A similar work~\cite{pinot2018} defines neighboring weight functions in terms of the $L_{\infty}$-norm and uses the exponential mechanism to release a private approximation of the minimum spanning tree topology for a given graph. 

\subsubsection{Distributed Private Query Release}
\label{sec:dist_local_dp}

So far, we have reviewed published works for releasing graph statistics in the edge and node differential privacy frameworks and some variants. One thing they have in common is a trustworthy centralized data curator who collects sensitive information from participants and answers analysts' queries in a private manner. This is also known as the \textit{centralized differential privacy} (CDP) model. 
In contrast, in the \textit{local differential privacy} (LDP) model \cite{Kasiviswanathan2011} there is no central data curator 
and each individual holds their sensitive information locally. When an analyst wants to calculate a global statistic over the population, each individual first answers the query locally in a private manner, and the analyst then collects the private local statistics from all the individuals and tries to get an aggregate view. 
Because it affords contributors strong privacy protection, local differential privacy has become a popular research topic in recent years. 

An earlier work~\cite{sun2019analyzing} proposes
mechanisms for releasing private triangle, three-hop path, and k-clique counts
in a localized setting called \textit{decentralized differential privacy} under the edge DP framework. 
In this new privacy model, each node shares its subgraph count in a private manner that not only protects its connected edges but also edges in its \textit{extended local view (ELV)}, i.e., the two-hop neighborhood. To protect its ELV, a node must add noise proportional to the counting-query sensitivity calculated from neighboring ELVs, which are ELVs of the nodes in neighboring global graphs. This definition of a neighboring ELV leads to large query sensitivity because two neighboring ELVs could have different sets of nodes. Thus, the authors present a two-phase framework to reduce noise magnitude using local sensitivity. The first phase determines an upper bound over all user local sensitivities. Each node in the second phase then shares its count privately according to this upper bound. The first phase includes two steps, which estimate the second order local sensitivities~\cite{karwa2011private} and derive an upper bound for the local sensitivities from that estimation.

Another work~\cite{ye2020} presents a \textit{local framework for graph with differentially private release} (LF-GDPR), which is claimed to be the first LDP-enabled graph metric estimation
framework for general graph analysis.
It designs an LDP solution
for a graph metric estimation task by local perturbation, collector/curator-side aggregation, and calibration. It makes the assumption that the target graph metrics can be derived from \emph{atomic metrics}, in particular the \emph{adjacency bit vector} and \emph{node
degree}. An optimal solution is then described for the allocation of privacy
budget between an adjacency bit vector (derived from a graph adjacency matrix) and node degree.
LF-GDPR is stated to enable solution generality and
estimation accuracy from LDP perturbation of these two atomic metrics, from which a further range of graph metrics can be derived.
The authors show how LF-GDPR can be used to compute clustering coefficients and perform community detection. 
%
%

\subsubsection{Private Graph Data Mining}
\label{sec:pmining}
In this section, we survey more complex graph queries that show how certain Graph Data Mining algorithms \cite{aggarwal10} can be done in a differentially private manner. 

\paragraph{Frequent Pattern Mining}
Frequent subgraph mining counts the number of occurrences of a 
given subgraph in a collection of graphs.
A subgraph is considered frequent if it occurs more 
than a user-specified threshold. The problem has wide applications in 
bioinformatics and social network analysis. While mining subgraphs of interests 
is an attractive practical problem, privacy concerns arise when the collection 
of graphs contain sensitive information.

The first differentially private subgraph mining mechanism \cite{shen2013mining} uses the exponential mechanism to release the top $k$ most frequent subgraphs. 
The main challenge of applying the exponential mechanism directly to obtain frequent 
subgraphs is calculating the normalizing constant in the sampling distribution because it
is infeasible to enumerate the output space. 
For this reason, the authors use the 
Metropolis-Hasting algorithm to sample frequent subgraphs from a Markov chain, 
in which each chain node is a subgraph and each chain edge represents an 
operation, which is one of edge addition, deletion and node addition. The 
proposed mechanism is shown to be $\varepsilon$-differentially private if the 
Markov chain converges to a stationary distribution. 


Another related mechanism~\cite{xu2016differentially} proposes to return frequent subgraphs with a different number of edges up to a maximal number. To release private subgraph
counts, it adds Laplace noise that is calculated using a lattice-based method to reduce
the noise magnitude. Since subgraphs with $m$ edges can be obtained from 
subgraphs with $m-1$ edges by adding one edge, this subgraph inclusion 
property can be used to create a lattice in which each lattice point is 
a subgraph. This lattice partitions a collection of graphs into mutually
disjoint subsets based on lattice paths of the 
frequent subgraphs. This technique eliminates irrelevant graphs from the domain,
thus reducing the sensitivity of subgraph-counting queries and the amount of noise for a given level of DP. 

\paragraph{Subgraph Discovery}
An earlier work~\cite{Kearns16} provides a different privacy model over social network 
data to identify a targeted group of individuals in a 
graph. This model is of particular interest in domains like criminal 
intelligence and healthcare. It provides privacy
guarantees for individuals that do not belong to the targeted group of interest. 
To achieve that, it introduces a graph-search algorithm that is 
based on a general notion of the proximity statistics, 
which measure how close a given individual is to the targeted set of 
individuals in the graph. 
This algorithm performs an iterative search over the graph for 
finding $k$ targeted disjoint connected components. In each iteration, 
the algorithm starts from a given node $v$ and finds the set of all nodes 
that are part of the same connected component as $v$. It
ensures the search for a new component is modified via randomization, which 
follows node DP by sampling noise from the Laplace 
distribution. The privacy cost of the algorithm increases with the 
number of targeted disjoint connected components (subgraphs defined 
on targeted individuals), and not with the total number of nodes 
examined. Thus, the privacy cost can be small if
the targeted individuals appear only in a small number of connected 
components in the graph. 

\paragraph{Clustering}
\citet{pinot2018} recently propose a method that combines 
a sanitizing mechanism (such as exponential mechanism) with a 
minimum spanning tree-based clustering algorithm. 
Their approach
provides an accurate method for clustering nodes in a graph while 
preserving edge DP. The proposed algorithm is able to recover arbitrarily shaped 
clusters based on the release of a private approximate minimum spanning
tree of the graph, by performing cuts iteratively to reveal the clusters.
At every iteration, it uses the exponential mechanism to find the next 
edge to be added to the current tree topology while keeping the weights 
private, which provides a trade-off between the degree of privacy and 
the accuracy of the clustering result.

\paragraph{Graph Embedding}
Graph embedding \cite{hamilton17} is a relatively new graph-analysis paradigm
that encodes the vertices of a graph into a low dimensional vector space
in a way that captures the structure of the graph. \cite{xu2018} 
studies the use of matrix factorization to achieve DP 
in graph embedding. The application of Laplace and exponential
mechanisms can incur high utility loss in existing random-walk-based 
embedding techniques because of the large amount of edge sampling required
and the sensitivity of stochastic gradients. Thus, that study proposes a
perturbed objective function for the matrix factorization, which achieves 
DP on the learned embedding representations. However, 
to bound the global sensitivity of the target non-private function, 
it requires complex analytic calculations that scale poorly. 
A following work~\cite{zhang2019} proposes to use a Lipschitz 
condition \cite{raskhodnikova2016lipschitz} on the objective function of
matrix factorization and a gradient clipping strategy to bound the global
sensitivity, with composite noise added in the gradient descent to 
guarantee privacy and enhance utility.

\tr{
\paragraph{Graph Neural Networks}
Graph Neural Networks (GNNs) \cite{zhou-survey-2020} are designed to improve the computational efficiency and generalization ability of graph embedding
techniques. GNNs have superior performance in learning node representations for various graph-inference tasks, including  node classification and missing-value imputation, edge or link prediction, 
and node clustering \cite{wu2020}. 
While the use of DP in traditional graph analysis and statistics 
applications is now reasonably well established, 
there are significantly fewer studies on differentially private GNN 
training methods~\cite{de2021}. 

A recent work~\cite{mueller2022} shows how the procedure of \textit{differentially 
private stochastic gradient descent} (DP-SGD) \cite{abadi2016} can be transferred from
database queries to multi-graph learning tasks where each graph can be 
seen as an individual entity in a multi-graph data set. However, this 
approach cannot be applied to GNNs in a single-graph setting  
because the individual data points (nodes or edges) in a graph cannot be
separated without breaking up the graph structure.

Another study~\cite{igamberdiev2021} proposes a random graph splitting 
method to graph convolutional networks, by partitioning a given graph into 
smaller batches to approximate 
sub-sampling amplification and then applying differentially private versions of 
gradient-based techniques like DP-SGD and Adam \cite{kingmaB14} for training.
This method provides higher training efficiency and privacy
amplification by sub-sampling. However, it results in tighter 
privacy bounds than when applied to the whole population. To address 
this problem, ~\cite{sajadmanesh2021} proposes to apply 
local DP~\cite{sun2019analyzing} on the node features, without protecting
the graph structure. Following the strategy of Private Aggregation of 
Teacher Ensembles (PATE)~\cite{papernot2016}, \cite{olatunji2021} 
recently proposes different teacher-student models to allow the differentially private 
release of GNNs.

Several recent works \cite{zhang2022,he2021,zhang2021} discuss the possibility
of performing privacy attacks against GNNs and quantify the privacy leakage of 
GNNs trained on sensitive graph data. For example, \cite{zhang2021} shows that 
DP in its canonical form cannot defend against a possible privacy attack 
while preserving utility.
Thus, more research is required to investigate how differential privacy noise can be
added to graphs to protect sensitive structural information against 
these privacy attacks.  

}

\section{Private Graph Releases}
\label{sec:syngraph}


Besides answering graph queries privately, the other popular way of protecting sensitive information in graph data is by generating synthetic graphs similar to the original ones. 
The major advantage of such methods is that they are independent of graph queries and hence can be used to answer any subsequent graph questions with low or no risks of privacy leakage. Recent works \cite{zhang2019utility,casas2020duef} propose generic strategies for evaluating the utility and privacy tradeoff in synthetic graphs to give guidance on how existing private mechanisms perform.   
In general, a private graph release mechanism can be evaluated on a variety of graph statistics to reflect its reconstruction accuracy from different perspectives. Such evaluations can be compared with other private mechanisms or with non-private graph-generation mechanisms to determine the impact on utility under different privacy settings.

This section surveys the main approaches that have been proposed for releasing synthetic graphs. 
\Cref{subsec:syn graphs non provable} touches on non-provable methods and refers readers to a few existing survey papers in that area. 
Our focus is principally on provable methods, which are discussed in detail in \Cref{subsec:syn graphs provable}. 


\subsection{Synthetic Graphs with No Provable Privacy Guarantee}
\label{subsec:syn graphs non provable}
There have been numerous studies on how to release a synthetic graph that is a close approximation to the original graph, while making it difficult to reconstruct the original graph or identify individuals in it. 
If one does not need any guarantee of the level of protection on the generated synthetic graphs, there are plenty of options for doing so, ranging from edge/vertex perturbation-based to sampling-based to generalization-based techniques and so on.  

The main problem with these methods is that they do not provide any mathematical
guarantees of privacy.
For example, consider $k$-anonymized graphs. 
In a $k$-anonymized data set, each record is indistinguishable from at least $k - 1$ other records with
respect to certain identifying attributes~\cite{sweeney2002k,samarati2001protecting}. A $k$-anonymized data set can have major privacy problems due to a lack of diversity of sensitive attributes. In particular, the degree of privacy protection does not depend on the size of the quasi-identifier attribute set, but is rather determined by the number of distinct sensitive
values associated with each quasi-identifier attribute set~\cite{machanavajjhala2007diversity}. Second, attackers
often have background knowledge, and $k$-anonymity does not guarantee privacy
against background knowledge attacks~\cite{machanavajjhala2007diversity}.
%
For more details and the milestone works that have been done in this line of research, the interested readers can read these surveys  \cite{chakrabarti2006,wu2010survey,casas2017survey,kiranmayi2020review}. 

\subsection{Synthetic Graphs with Provable Privacy Guarantee}
\label{subsec:syn graphs provable}
The previous Section described private graph release mechanisms that rely on the amount of background information adversaries have about the sensitive data. As this is difficult to anticipate, 
they do not provide strong privacy guarantees.
This Section reviews alternate approaches, which are based on the
edge differential privacy notion, and thus provide provable privacy guarantees.

\subsubsection{Generative Graph Models}

Before privacy was addressed in public graph data, researchers had been working on generative graph models to replicate an underlying unknown data-generating process. These models usually have parameters that can be estimated from a given class of graphs. Thus, one way to synthesize a provably private graph is to ensure the parameters of such generative models are estimated in a way that is differentially private. For this reason, there is a close connection between private query-answering mechanisms and private synthetic-graph generation mechanisms. 

\emph{Pygmalion} \cite{sala2011sharing} is an example of such mechanisms, where a $dK$-graph model \cite{mahadevan2006} is used to capture the number of connected $k$-node subgraphs with different degree combinations into $dK$-series.
These $dK$-series are then sorted and partitioned into disjoint unions of \emph{close} sub-series, each of which is made private by the Laplace mechanism. A further noise reduction to the entire series is performed using the constraint inference method \cite{hay2009accurate}.
Pygmalion is applied to three real graphs with tens to hundreds of thousands of nodes and tested under some popular graph metrics (e.g., degree distribution, assortativity, graph diameter, etc.) and two application-level tasks, spam filter and influencer identification.  It shows a limited impact on the generated synthetic graphs across a range of privacy budgets when compared with its non-private alternatives. 

A following work \cite{wang2013preserving} improves the utility of the $dK$-graph model method by adding Laplace noise proportional to smooth sensitivity (\Cref{equation:smooth sensitivity}). 
The test subjects, i.e., $dK$-1 and $dK$-2 based mechanisms, outperformed a non-private \textit{stochastic Kronecker graph} (SKG) generator 
on four real networks as demonstrated in \cite{wang2013preserving}. 
Although $dk$-2 has higher utility than $dk$-1, it is only superior for very large privacy budgets. 
A more recent work \cite{iftikhar2020dk} reduces the noise magnitude by adding a microaggregation step to the $dK$-series before adding Laplace noise. This microaggregation step partitions $dk$-series into clusters of similar series and replaces each cluster with a cluster prototype, providing an aggregated series with lower sensitivity. 

Another method \cite{mir2012differentially} uses an SKG model \cite{leskovec2007scalable} that recursively creates self-similar graphs by using Kronecker product of the adjacent matrix of an initiator graph to itself. 
The adjacency matrix entries of an initiator graph are the estimated SKG model parameters from a given graph. 
Following \cite{gleich2012moment}, four different subgraph counts (i.e. edges, triangles, 2-stars, 3-stars)
are selected as the SKG parameters.
The number of triangles is calculated using smooth-sensitivity-based Laplace noise.
The other three subgraphs are counted from a private degree sequence produced by the constraint inference algorithm \cite{hay2009accurate}. 
This private SKG-based mechanism produces synthetic graphs with comparable graph statistics as the ones
produced by two non-private SKG models. 

A different approach in \cite{xiao2014differentially} utilizes the \textit{Hierarchical Random Graph} (HRG) model \cite{clauset2006structural} to encode a network in terms of its edge probabilities. 
An HRG model of a given graph consists of a dendrogram and the associated probabilities. The dendrogram is a rooted binary tree, where each leaf node corresponds to a node in the given graph and each internal node has an associated probability. 
The probability that two nodes are connected in the original graph is captured by the probability of their lowest common ancestor in the dendrogram.
The authors use a \emph{Markov chain Monte Carlo} method (MCMC) to select a good dendrogram. 
This MCMC samples through the space by varying the subtree rooted at a randomly picked internal node from the dendrogram. 
This step also ensures dendrogram privacy, as the MCMC plays a similar role as the exponential mechanism. 
This HRG-based mechanism shows superior performance over the $dK$-2-based mechanism \cite{wang2013preserving} and a spectral method \cite{wang2013differential} when applied to four real networks.

A mechanism that releases private graphs with node attributes is \cite{jorgensen2016publishing}, which is based on the \textit{Attributed Graph Model} (AGM)\cite{pfeiffer2014attributed}. 
AGM has three classes of model parameters, each of which can be privately estimated. First, node attribute parameters are estimated using counting queries with the Laplace mechanism due to their low sensitivity. 
Second, attribute-edge correlation parameters 
are also estimated using counting queries, but on a projection \cite{blocki2013differentially}
of the original graph into a bounded maximum degree one.
Third, edge-generation parameters are modeled by a generative model, called \textit{TriCycLe}, which simulates the degree sequence and clustering coefficients. This model is used with two parameters, the degree sequence and triangle count, both of which are then privately estimated by methods from \cite{hay2009accurate} and \cite{zhang2015private}, respectively. 

\subsubsection{Graph Matrix Perturbations}
Besides generative graph models, another approach is to release private approximations of the original adjacency and Laplacian matrices using matrix perturbation strategies. 

Examples of such mechanisms are proposed in~\cite{wang2013differential} to release the largest $k$ eigenvalues $\{\lambda_1, \dots, \lambda_k\}$ and eigenvectors $\{u_1, \dots, u_k\}$ of an adjacency matrix, which can be turned into a lower rank adjacency matrix by $M_A^k = \sum_{i=1}^k \lambda_i u_i u_i^T$. 
A first mechanism directly adds Laplace noise to the top $k$ eigen pairs in proportion to their global sensitivity. 
A second mechanism uses a Gibbs sampler \cite{hoff2009simulation} to sample the largest $k$ eigenvectors from the matrix \textit{Bingham-von Mises-Fisher} (BMF) distribution, which is a probability distribution over orthonormal matrices, such as the eigenvector matrix.
The first mechanism outperforms the second one in many experimental settings on real network data.

Another contribution~\cite{chen2014correlated} focuses on generating private graphs from original
graph data in which the existence of an edge may be correlated with the existence of other ones.
It introduces an extra new parameter $k$ that controls the maximum number of correlated edges and evenly splits the privacy budget $\epsilon$ to $\epsilon/k$.
This setting is consistent with the $k$-edge differential privacy framework \cite{hay2009accurate}.
Once the privacy budget is evenly split, the adjacency matrices are sanitized as if they were not correlated. 
The adjacency matrix perturbation process contains node relabeling, dense region discovery and edge reconstruction using the exponential mechanism. 
These first two steps find high density regions in the adjacency matrix, which can then be reconstructed with accuracy~\cite{gupta2012iterative}. Each of these three steps receives a portion of $\epsilon/k$ as its privacy budget. This proposed \textit{density-based exploration and reconstruction} (DER) outperforms a simple Laplace mechanism and a variation of DER random edge reconstruction in all test cases~\cite{chen2014correlated}. 

A drawback of the HRG \cite{xiao2014differentially} and the DER \cite{chen2014correlated} mechanisms is their quadratic running time in the number of nodes. A more efficient matrix perturbation mechanism is \textit{Top-m Filter (TmF)} \cite{nguyen2015differentially}, which runs linearly in the number of edges. It starts by adding Laplace noise to each cell in the adjacency matrix, then only chooses the top noisy cells as edges in the perturbed matrix. 

In \cite{brunet2016novel}, the authors focus on releasing private adjacency matrices for weighted directed graphs, and define different neighboring graphs than in edge and node DP. They use the Laplace mechanism to add noise to the adjacency matrices according to the sensitivity within blocks of entries, as some edge weights are less sensitive than others and so should be treated differently. They then propose an automated method to partition the matrix entries without any prior knowledge of the graph.

Although not intended to deal with graph data, some earlier works on publishing low-rank private approximations of matrices could be used to publish sanitized private adjacency or Laplacian matrices.
For example, in \cite{blum2005practical} the authors study the connection between singular value decomposition and eigen decomposition of matrices. More precisely, given a matrix $M_A$, they used the top $k$ eigenvalues of a perturbed matrix $M_A\cdot M_A^T$ with carefully calculated Gaussian noise to privately approximate $M_A$. The truncated eigenvalues can then be used to obtain a rank-$k$ approximation of the original matrix. A subsequent work improves on the utility of the approximated low-rank matrices~\cite{hardt2011beating}, but that is inapplicable to graph data due to the unbalanced dimension constraint on the input matrices to satisfy a low-coherent assumption. Although proved under the assumption that the input matrices are symmetric and positive semidefinite, the work in \cite{kapralov2013differentially} can be generalized to symmetric matrices and still improves the quality of the released low-rank matrices under $\epsilon$-differential privacy. The strategy is to use the exponential mechanism to sample a rank-1 approximation of the given matrix $M_A$ with the utility function being proportional to $exp(z^T\cdot M_A\cdot z)$, hence this requires $M_A$ to be positive semidefinite. 
This step is repeated $k$ times and the sampled rank-1 vector $v_i$ is accumulated in the form 
$v_i^T\cdot M_A\cdot v_i+Lap(\frac{k}{\epsilon})$ to get a final private rank-$k$ approximation of $M_A$. 

\subsubsection{Distributed Private Graph Release}
\label{subsubsection:Distributed Private Graph Release}
%
%
In the above approaches, a single data custodian
knows about the entire input graph, then applies a DP-based mechanism to 
release a synthetic version of it.
As we discussed in \Cref{sec:dist_local_dp}, local 
differential privacy (LDP) can be used to protect sensitive information of 
individuals from an untrustworthy data custodian~\cite{Kasiviswanathan2011}. 
In this setting, each data source locally perturbs sensitive data
before sending them to the data curator to construct a representative graph. 

LDPGen~\cite{qin2017} is one of the first LDP-based private synthetic
graph generation techniques. It groups nodes with similar degree vectors 
into the same cluster. 
Inter and intra-cluster edges can then be 
generated to get a private synthetic graph. More specifically, the clustering
step starts with a random node clustering. Each node $u$ then shares a noisy
degree vector $(\Tilde{\sigma}_1^u, \dots, \Tilde{\sigma}_{k_0}^u)$ under 
the current clustering scheme, where $\sigma_1^u$ is the degree of $u$ in 
the first cluster.
Each noisy degree vector satisfies the LDP property.
The data curator updates the clustering
scheme once all noisy degree vectors are received. The updated clustering 
scheme is communicated back to all individuals for refinement of their 
private degree vectors, who then share with the data curator again for a 
further update round. 
When compared against two simpler mechanisms~\cite{qin2017}, LDPGen generates
synthetic graphs with higher utility under different use cases,
such as community discovery.

Another LDP-based approach focused on preserving the structural utility of the 
original graph $G$ \cite{gao2018local}.
To generate the synthetic graph, the authors first split $G$ into multiple subgraphs.
They then carefully select a set of subgraphs without any mutual influence to be sanitized.
They use the
\emph{hierarchical random graph} (HRG) model~\cite{clauset2006structural} 
to capture the
local features from each of the selected subgraphs. LDP is then introduced
into each HRG such that the corresponding subgraph in $G$ is regenerated
according to the sanitized HRG.
This produces an updated privatized graph $G$. 
The added local noise on each HRG preserves more structural 
information compared to applying global differential privacy on the 
original graph. It provides synthetic graphs that more closely match
the original ones, thus providing higher utility.

\subsubsection{Iterative Refinement}
Some strategies iteratively look for the best synthesized graph data that is both private and close to the original graph under the guidance of an objective function. 
For example, in the IDC mechanism described in \cite{gupta2012iterative} removes bad approximated graphs until a good candidate is found.
In \cite{gupta2012iterative}, the graph data consists of the weighted edges between pairs of nodes and the synthetic graph-release mechanism for linear queries (including cut queries) is built upon the Laplace mechanism to add noise to edge weights, but with an additional linear programming step that solves for a close approximation where weights are restricted to $[0,1]$ to remove negative edge weights. 
In \cite{proserpio2012workflow,proserpio2014calibrating}, an MCMC-based mechanism starts from a random graph generated from a differentially private degree sequence then searches for an optimal graph with edge swapping operations to better fit the weighted Privacy Integrated Query (wPINQ)  measurements, but in a way that remains consistent with the given private degree sequence. 


\section{Beyond Differential Privacy: Limitations and Alternatives}
\label{sec:limits}

As we have seen throughout this paper, differential privacy is by far the most 
popular framework for analyzing and designing provably private graph data release 
algorithms. In this section, we discuss some known limitations of the
differential privacy framework, especially as applied to graph data, and 
describe other formal privacy definitions as alternative frameworks.

\subsection{Differential Privacy on Correlated Data}
\citet{kifer2011no} discuss how 
correlation among records in a data set can affect the privacy
guarantees of DP mechanisms. In particular, they
point out that the real-world concept of privacy often cannot be
modeled properly using only the existence or otherwise of an 
entity's record in the data, but needs to take into account the 
{\em participation} or otherwise of an entity in the data-generating 
process. To illustrate this, the authors give examples of social 
network data where, given two graphs $G_1$ and $G_2$ that differ 
only in one edge and a randomized algorithm that evolves $G_1$ 
into $G_1'$ and $G_2$ into $G_2'$ using a common model like the 
Forest Fire model \cite{leskovecKF07}, the query on the number of 
edges between two communities in the evolved graph, at best, cannot
be answered in a differentially private way with sufficient utility
and, at worst, is answered in a way vulnerable to attacks because 
the underlying privacy issue is modeled incorrectly as a single edge
difference between $G_1'$ and $G_2'$ instead of that on $G_1$ and
$G_2$. The basic observation is that path dependency in a 
graph-evolution model like Forest Fire results in correlated 
data in $G_1'$ and $G_2'$ that allow their origins $G_1$ and $G_2$ 
to be easily distinguished.

More generally it is claimed in~\cite{kifer2011no} that under almost
any reasonable formalization, the assumption that evidence of 
participation can be encapsulated by exactly one record is implied by 
the assumption that all the records are generated independently, although
not necessarily from the same distribution. This is subsequently formally 
proved in \cite{kifer14}. From that perspective, the applicability of 
DP is limited by that independence assumption.
This is a serious limitation because real-world data, especially 
graph data, are often complex and exhibit strong correlations 
between records. These correlations between records exist due to 
behavioral, social, or genetic relationships between 
users~\cite{liuMC16,zhao2017}. For example, in social network data, 
it is highly likely that the locations of friends exhibit strong 
correlations since they tend to visit the same places.

%
%
Following the theoretical work in \cite{kifer2011no}, \citet{liuMC16} use an inference attack to demonstrate 
the vulnerability of applying differential privacy mechanisms on correlated 
graph data.
In their experiments on social network data, they show that an 
adversary can infer sensitive information about a user from private query 
outputs by exploiting her social relationships with other users that share
similar interests. Such social and behavioral correlations between users 
have also been used to perform de-anonymization attacks on released 
statistical data sets~\cite{srivatsa2012}.  
In another related work~\cite{almadhoun2020} the authors show
that an adversary has the ability to infer information about 
an individual in a statistical genomic data set using the information 
of their related other household members. For example, they can infer the 
susceptibility of an individual to a contagious disease by using the 
correlation between genomes of family members. 

\subsection{Dependent Differential Privacy}

The notion of \emph{dependent differential privacy} (DDP)~\cite{liuMC16}
considers correlations between records in a statistical data set to overcome
inference attack by adversaries who have prior information about the probabilistic
dependence between these records.
DDP introduces the novel concept of  
\emph{dependence coefficient} that quantifies the level of correlation 
between two records.
Summing the dependence coefficients between a record and all 
other records correlated with it, we can get a quantification of how 
changes in each record can affect other related records in a data set.
The maximum dependence coefficient allows a user to calculate the dependent 
sensitivity for answering the query over a correlated data set. Thus, this
sensitivity measure can then be used to instantiate the Laplace mechanism 
to achieve privacy while minimizing noise. However, in practice, the 
effectiveness of the DDP framework on tabular data is limited by how well 
the correlation among data can be modeled 
which is a challenging problem in itself.


In contrast, graphs show correlations that are inherent among nodes as 
relationships between nodes represent how the nodes are connected. 
Assuming each node has a degree of $m$, a modification of an attribute
value of a node potentially causes changes in at most $m-1$ other nodes
due to the probabilistic dependence relationships between nodes. Thus, 
the dependence coefficient measure can be used in graph data
to quantify the amount of noise that needs to be 
added to each query answer considering the nodes and their surrounding
neighbors in the graph. As an example, in~\cite{zhao2017} the authors
use probabilistic graphical models to explicitly represent the dependency 
between records, and show how the structure of correlated data can be 
carefully exploited to introduce noise into query responses to achieve
higher utility.

\subsection{Pufferfish Privacy}

%

The above works on correlated data have led to further studies on 
generalizing DP as a framework that can be customized to 
the needs of a given application. We now discuss one such
framework called \emph{Pufferfish}~\cite{kifer14} that 
makes it easier to 
generate new privacy definitions with rigorous statistical guarantees about 
the leakage of sensitive information. Building on Pufferfish, 
several classes of privacy definitions have recently been proposed, including
Blowfish \cite{HeMD13}, and 
Bayesian DP~\cite{yang2015bayesian}.


The Pufferfish framework defines by the following components: 
a set $\mathcal{S}$ of potential secrets, 
a set $\mathcal{S}_{\text{pairs}} \subseteq \mathcal{S} \times \mathcal{S}$ of 
mutually exclusive pairs of secrets to be protected, and  
a set $\mathcal{D}$ of data-generation processes.  
The set $\mathcal{S}$ serves as an explicit specification of what we would
like to protect, for example, the record of an entity $x$ is/is not in the data set. 
The set $\mathcal{S}_{\text{pairs}}$ represents all the pairs of secrets 
that should remain indistinguishable from each other given the query response.
Finally, $\mathcal{D}$ represents a set of assumptions about how the data 
evolved (or was generated) that reflects the adversary’s belief about 
the data, for example probability distributions, variable correlations, 
and so on. 

\begin{definition}[\cite{kifer14}]
\label{def:pufferfish}
Given $\mathcal{S}$, $\mathcal{S}_{\text{pairs}}$, $\mathcal{D}$, and a privacy 
parameter $\epsilon > 0$, a randomized algorithm $\mathcal{M}$ satisfies 
$\epsilon$-$\mathbf{Pufferfish}(\mathcal{S}, \mathcal{S}_{\text{pairs}}, 
\mathcal{D})$ privacy if $\forall\Om\in \mathit{range}(\mathcal{M})$, 
$\forall (s_i,s_j) \in \mathcal{S}_{\text{pairs}}$, 
$\forall \theta \in \mathcal{D}$, and for each data set $D$ that can be 
generated from $\theta$, we have
\begin{gather}
    Pr[\mathcal{M}(D) = \Om \,|\, s_i,\theta] \leq e^\epsilon Pr[\mathcal{M}(D) = 
    \Om \,|\, s_j,\theta] \label{eqn:pufferfish1} \\
    Pr[\mathcal{M}(D) = \Om \,|\, s_j,\theta] \leq e^\epsilon Pr[\mathcal{M}(D) = 
    \Om \,|\, s_i,\theta] \label{eqn:pufferfish2},
\end{gather}
where the probabilities are taken over the randomness in $\theta$ and $\mathcal{M}$.
\end{definition}

One can show that the inequalities (\Cref{eqn:pufferfish1}) and 
(\Cref{eqn:pufferfish2}) are equivalent to the following condition on the odds
ratio of $s_i$ and $s_j$ before and after seeing the query output $\Om$: 
\begin{equation}\label{eqn:pufferfish_odds_ratio}
    e^{-\epsilon} \leq \frac{Pr[s_i \,|\, \mathcal{M}(D) = \Om, \theta] / 
    Pr[s_j \,|\, \mathcal{M}(D) = \Om, \theta]}{ Pr[s_i \,|\, \theta] / 
    Pr[s_j \,|\, \theta]} \leq e^\epsilon.
\end{equation}

Recall that in the Pufferfish framework, each probability distribution 
$\theta \in \mathcal{D}$ corresponds to an attacker's probabilistic beliefs 
and background knowledge. Thus, for small values of $\epsilon$, 
\Cref{eqn:pufferfish_odds_ratio} denotes that observing the query output
$\Om$ provides little to no information gain to attackers $\theta$ who are trying
to infer whether $s_i$ or $s_j$ is true. When we assume each record in the 
data set is independent of one another such that no correlation exists, then 
the privacy definition of Pufferfish is the same as the privacy definition 
of $\epsilon$-differential privacy. 

To the best of our knowledge, the only mechanism that currently provides
Pufferfish privacy is the Wasserstein mechanism~\cite{song2017}, which is a
generalization of the Laplace mechanism for DP. In~\cite{song2017}, the authors
prove that it 
always adds less noise than 
other lesser used mechanisms such as
group differential privacy~\cite{palanisamy2017group}.
This makes Pufferfish-based techniques more applicable to graph data. 
For example, let us assume a connected graph $G=(V,E)$ where each node $v\in V$ 
represents an individual and each edge $(v_i,v_j)\in E$ represents the relationship
between two individuals. The set of edges can be interpreted as values in the domain 
that an adversary must not distinguish between; i.e., the set of discriminative
secrets is $S^G_{pairs}=\{(s_{v_i},s_{v_j}): \forall(v_i,v_j)\in E\}$.
Following \Cref{def:pufferfish}, one can add Laplacian noise equal to 
$(\Om/\epsilon)$ to achieve $\epsilon$-Pufferfish$(V\times V, S^G_{pairs}, G)$ 
privacy for any $V'\subseteq V$, where $w$ is the average size of a connected
component in $G$.

\subsection{Other Provable Privacy Definitions}

Inferential Privacy~\cite{ghosh2017} is a similar notion to Pufferfish privacy.
It relies on modeling correlated data as a Markov Chain, and adding noise proportional
to a parameter that measures the correlation. Its 
mechanisms are less general than the 
Wasserstein mechanism~\cite{song2017}, but are applicable to a broader
class of models than the Markov quilt mechanisms 
that measure the difference between an adversary’s belief about sensitive
inferences before and after observing any released data.


Other notions such as Adversarial Privacy~\cite{rastogi2009relationship} are
weaker than DP, but give higher utility when querying social networks under
certain assumptions. 
Adversarial privacy is achieved in graph queries if the prior and posterior 
of a data point after seeing the query output are almost indistinguishable. In~\cite{rastogi2009relationship}, the authors 
restrict adversaries' prior distributions to a special class of 
distributions and prove that adversarial privacy 
is equivalent to $\epsilon$-indistinguishability, a generalization of 
$\epsilon$-differential privacy, in the sense that neighboring data sets 
$D_1$ and $D_2$ satisfy either 
$|D_1|=|D_2|=n$ and $|D_1 \oplus D_2|=2$ 
or $|D_1|=|D_2|+1$ and $D_2\subsetneq D_1$.
They further state
that adversarial privacy
can be applied to stable queries, such as subgraph counting queries. 


Compared to the above, Zero-Knowledge Privacy \cite{gehrke2011towards}
proposes a stronger privacy definition to protect 
individual privacy where differential privacy may fail. It argues
that the standard concept of differential privacy may not protect 
individual privacy in some social networks, where specific auxiliary
information about an individual may be known. Given an adversary who
has access to a mechanism that runs on a data set, Zero-Knowledge 
Privacy says that the outputs gained by an adversary with or without 
accessing the mechanism with aggregated information will be similar.
The choice of aggregate information is sensitive to the privacy 
concept, such as the use of aggregate
information from any computation on a subset of randomly chosen 
$k$ samples from the given data set with an individual's record 
concealed. With this setting, Zero-Knowledge Privacy satisfies 
composition and group privacy just as differential privacy does. 
Using an extended formulation for graphs with bounded degrees,
\citet{gehrke2011towards} proves Zero-Knowledge Privacy for the average-degree query and graph-edit-distance queries (in terms of edge addition
and deletion), as well as boolean queries like 
whether a graph is connected, 
Eulerian, 
or acyclic.

%


\section{Application Domains and Example Use Cases}
\label{sec:app}

The previous sections provided a domain-agnostic description
and classification of mechanisms to release graph data and answer graph queries
with enhanced privacy.
In contrast, this section discusses these same mechanisms in the context of
domain applications and industry sectors where private graph
analytics are used in practice to provide values. \Cref{tab:app}  summarizes
the examples of application domains discussed in this section and the related
surveyed contributions.
Moreover, some recent technology such as the Internet-of-Things (IoT) can be
applied to many of these domains, and produce data
that are best captured and analyzed using graph-based models and analytics.
Indeed, distributed IoT devices readily map to graph vertices, while their
relationships, interactions, and measured data map to graph edges.
Section 6.5 will also briefly introduce use-cases of IoT technology in the 
context of graph analytics for different domain applications.

\begin{table}[!htbp]
\footnotesize
\centering
\begin{tabular}{|l|l|} 
\hline
Application Domains & Example Graph Analytics (private and non-private) \\ 
\hline
\hline
Social Networks & \cite{rastogi2009relationship,zhu2017differentially,narayanan2009anonymizing,task2012guide,isaak2018,ji2019,tian2021} \\ 
\hline
Financial Services & \cite{weber2018a,pourhabibi2020,chandola2009,hay2009accurate,karwa2012differentially,proserpio2014calibrating,chakrabarti2006,rehman2012,shen2013mining,xu2016differentially,Kearns16,spirtes2000,sarfraz2019} \\ 
\hline
Supply Chains & \cite{wagner2010,tan2019,soni2014,neo4j1,neo4j2,guo2018,benvcic2019, ogbuke2022} \\ 
\hline
Health & \cite{pfohl2019,leskovec2007,Dankar2013,kim2018,day2016publishing,mohan2012,nissim2007smooth,suriyakumar2021,palanisamy2017group,KiferM12,srivastava2019, sharma2018} \\
\hline
\end{tabular}
\caption{\label{tab:app}Example of application domains of graph analytics with some related surveyed contributions.}
\vspace*{-2em}
\end{table}

\subsection{Social Networks and Related Services}

Over the past few decades, social networks have become a global platform for 
individuals to connect with each other. These platforms allow 
third party businesses and their advertising partners to access an 
unprecedented level of information, which can be used to reach potential 
customers, or for better social targeting~\cite{kiranmayi2020review}. 
However, the sensitive information of individuals contained in social networks
could be leaked due to insecure data sharing~\cite{abawajy2016}.
For example, in early 2018, it was reported that up to 87 million Facebook 
users’ personal information and their social relationships might 
have been shared with a political consulting company, Cambridge 
Analytica, without individuals authorizations~\cite{isaak2018}. 
Differentially private
techniques can be used to ensure the privacy of individuals is preserved 
when social network data is queried or 
published~\cite{narayanan2009anonymizing,rastogi2009relationship, zhu2017differentially}. 

One real-life application of graph analytics to social networks is
described in~\cite{saqr2018}, where the authors use graph metrics on social
interactions between 82 students and lecturers in the context of courses at
Qassim University. They identify patterns associated with learning
outcomes, and  use these graph-derived insights to design and 
apply interventions to improve student engagement, marks, and
knowledge acquisition.
For example, they use in/out degree sequence metrics to estimate students' level
of activity, or closeness and betweenness centrality metrics to identify roles in
collaborating groups of students.
As surveyed earlier in this paper, many provably private mechanisms have 
been proposed for these specific graph metrics~\cite{kasiviswanathan2013analyzing,
day2016publishing,raskhodnikova2016lipschitz,roohi2019differentially}.

Community detection is
another common graph analysis, which is routinely performed on social network data,
to mine complex topologies and understand the relationships and interactions
between individuals and groups~\cite{task2012guide}.
For example in~\cite{lewis2008}, a team Social Science researchers performs such analytics
on graph data from students of a well-known U.S. college, to identify and
study ethnic and cultural communities and derives several structural and longitudinal
insights on them (e.g., taste in given music genre are more commonly shared within
a given ethnic group, etc.).
%
As illustrated by that real-world study, third-parties usually conduct such
community detection research, which could lead to significant privacy
breaches~\cite{isaak2018}.
%
\citet{ji2019} proposes a community detection algorithm, which protects the privacy of network
topology and node attributes. 
It formulates community detection as a maximum log-likelihood
problem that is decomposed into a set of convex sub-problems
on the relationships and attributes of one individual in the network.
It then achieves DP by adding generated noises to
the objective function of these sub-problems.

\subsection{Financial Services}
The financial sector is one of the earliest to embrace the 
{\em Big Data} revolution of the past decades~\cite{srivastava2015}. While
initial applications focused on extracting insights on internal data within
a single organization to provide added benefits to their stakeholders and
customers, more recent applications aim at data across businesses, types of
institutions, and countries, to produce further utility and 
financial benefits.
Graph analytics are the natural tools to derive knowledge from the networks of
data points, which arise when combining diverse data sets from multiple
sources.

The financial sector has been forecast to spend more than $\$$9 billion
annually to combat fraud~\cite{juniper2017}. Several graph-based techniques
have been proposed and surveyed to tackle this challenge~\cite{pourhabibi2020}.
In a specific real-world example~\cite{weber2018a}, forensic analysis researchers apply
several graph metrics, such as cycle detection, degree distribution, and PageRank, 
on graph data collected as part of the so-called Know-Your-Customer (KYC) process
of a private bank. Using these graph analytics, they achieve a reduction of $20-30\%$
on false-positive for the detection of suspicious financial activities by expert analysts.
However, these approaches often do not consider the privacy and confidentiality issues
that constrain the sharing of data between different financial organizations.
Many of the previously surveyed works provide privacy-enhanced alternatives
to these techniques. For example, some methods exploit the
density of groups of nodes and their interconnections~\cite{chandola2009},
which can be obtained through provably private degree sequences,
as described in
\cite{hay2009accurate,karwa2012differentially,proserpio2014calibrating}.


Modeling customer behaviors in financial services is another application
of graph analytics in the finance sector. It allows businesses
such as banks and insurers to better understand their customers, leading to
tailored products and services~\cite{beckett2000}.
In one existing trial~\cite{hadji2018}, data analysts use a credit scoring prediction solution, 
which model SMEs' financial behaviors into graph, and apply a process
based on adjacency matrices, correlation distances, node degree and closeness to
compute their credit scores. They partner with a European FinTech registered as a
Credit Rating Agency and apply this solution to their data. This produces 
credit scoring models that are significant predictors of loan defaults.
Identifying and
characterizing patterns in financial graph data
is another type of techniques to achieve such models of customer
behaviors~\cite{chakrabarti2006,rehman2012}. Several provably private methods
were previously described in \Cref{sec:pmining} to allow this pattern mining
\cite{shen2013mining,xu2016differentially,Kearns16,spirtes2000}.

\subsection{Supply Chain}
Supply chains are fundamentally graphs. Indeed, regardless of the sector, they
capture participants (e.g., food producer, manufacturer, transporter, retailer,
etc.) and their relationships (e.g., transactions, raw material supply/delivery,
subcontracting services, etc.), which form the nodes and edges of the corresponding
graphs. Thus it is not surprising that various graph analytics have been applied
to different types of supply chains for many purposes. 

Resilience is critical in supply chains, and several
graph analytics contributions focus on addressing this concern. 
In a recent industrial use-case~\cite{Hong2022}, researchers at Ford Motor Company propose a 
graph model to capture the flows and relationships of materials from suppliers to
finished products in their automotive supply chains. They further develop a novel
graph-based metric called Time-to-Stockout (TTS), which allows them to estimate the 
resilience of parts of these chains with respect to both market-side demand and
supply-side inventory. They apply their graph model and metric to real industrial
data from the Ford supply chain to demonstrate their effectiveness.
Other more traditional graph metrics, such as 
adjacency matrices can be used to identify and update the weakest parts in a given chain to increase its robustness~\cite{wagner2010}. Specific graph models
also exist to assess and increase structural redundancy in supply
chains~\cite{tan2019}. One example of such novel models defines a supply chain
resilience index (SCRI) to quantify resilience based on major structural enablers
in the graph and their interrelationships~\cite{soni2014}. In another example,
an approach based on node-degree is proposed to evaluate the resilience of
different Australian-based supply chains (e.g., lobster  and prawn fisheries,
iron ore mining)~\cite{tan2019}.

Optimization is another applied area of graph analytics in supply chain. In some
real-world use cases, the graph analytics provided by the Neo4J tool have been
used to lower response time in product quality management where several suppliers
are involved~\cite{neo4j1}, and to lower cost and complexity in inventory, payment and 
delivery management~\cite{neo4j2}.
In another case study on a laptop manufacturing supply chain, researchers use
a novel graph-based cost function and an algorithm based on similarity measures
to optimize the 
reconfiguration of a supply chain in order to lower the overall manufacturing
cost of the laptops~\cite{guo2018}.

\subsection{Health}

With the recent COVID-19 pandemic,
the healthcare industry has been developing
various systems that can mine insights from healthcare data to support
diagnoses, predictions, and treatments~\cite{alguliyev2021}.
Graph analytics allow researchers to effectively and efficiently 
process large, connected data.
However, due to the highly sensitive nature of patient data,
such data requires strong privacy guarantees before it can be released or 
used in healthcare applications. 

Differential privacy has been proposed as a possible approach to
allow the release of healthcare data with sufficient guarantee
against possible privacy attacks~\cite{pfohl2019}. For example, as we reviewed
in \Cref{sec:nodepriv}, node differential privacy can be applied to select 
nodes in a patient graph to detect outbreaks of diseases~\cite{leskovec2007}. 
Next, we discuss a few applications where graph-based differential
privacy techniques have been used in the healthcare domain. 

Along with the development of internet-of-things (IoT) technologies, 
smart healthcare services are receiving significant attention. They focus on
disease prevention by continuously monitoring a person’s health and providing
real-time customized services.
Such devices enable the collection of a vast amount of personal health data, which can be 
modeled as graphs. However, such graph data needs to be
treated with appropriate privacy techniques before they are used in data 
analytic processes as individuals can be re-identified by tracking and analyzing
their health data~\cite{Dankar2013}. 

An earlier work~\cite{kim2018} proposes a novel local differential privacy 
mechanism for releasing health data collected by wearable devices.
The proposed approach first identifies a small number of 
important data points from an entire data stream, perturbs
these points under local differential privacy, and then 
reports the perturbed data to a data analyst, instead of reporting all the graph data.
Compared to other approaches that release private graph data in the form of 
histograms, such as \cite{day2016publishing}, local differential privacy-based
approaches provide significant improvement in utility while preserving 
privacy against possible attacks~\cite{almadhoun2020}. Further, the development
of differentially private data analysis systems, such as GUPT~\cite{mohan2012},
allow efficient allocation of different levels of privacy for different user 
queries following the smooth sensitivity principle~\cite{nissim2007smooth}.


In another recent contribution~\cite{suriyakumar2021}, the authors analyze
different challenges present in healthcare data analysis. In their analysis,
they argue that different algorithms should be used to approximate group 
influences to understand the privacy fairness trade-offs in graph data. Thus,
some of the techniques discussed in \Cref{sec:limits}
can potentially result
in the high influence of majority groups in data sets as opposed to minority ones, 
imposing asymmetric valuation of data by the analysis model. This requires novel
privacy techniques, such as Pufferfish privacy~\cite{KiferM12}, to be used in 
such clinical settings to make sure minority class memberships are represented 
appropriately in data analysis models without compromising privacy.

\subsection{IoT Technology}

%
%
%

The Internet of Things (IoT) is a cross-cutting technology to many application domains
of graph analytics. This section presents some examples of the application of IoT in the 
context of private graph analytics for the previously discussed domains.

Social Internet of Things (SIoT) technology refers to IoT being deployed in social networks.
As an example, a recent work~\cite{tian2021} proposes a graph clustering privacy-preserving
method, which is based on structure entropy and combines data mining with structural
information theory. Through theoretical analysis and experimental evaluations, the authors show that their
privacy-enhanced clustering scheme provides a better utility/privacy trade-off than other 
schemes. 

Privacy-preserving applications for graph data have also been provided for the financial services domain. In ~\cite{sarfraz2019}, a privacy-aware IOTA ledger is presented for decentralized mixing and unlinkable IOTA transactions\footnote{IoTA is an open-source distributed ledger and cryptocurrency designed for the Internet of things (IoT)}. In ~\cite{sarfraz2019} the authors demonstrate the provision of privacy and security through a novel decentralized mixing protocol for the IOTA ledger, which incorporates a combination of decryption mixnets and multi-signatures.

In the supply chain sector, the distributed ledger (DL)-Tags~\cite{benvcic2019}
is an example of a solution that allows private graph analysis on distributed ledgers.
DL-Tags provides decentralized, privacy-preserving, and verifiable management of Smart
Tags through a product’s lifecycle. It is agnostic to the type of distributed ledger
being used, and provides evidence of the product’s origin and its journey across the
supply chain, while preventing tag duplication and manipulation.
A recently published review~\cite{ogbuke2022} on Big Data Supply Chain Analytics (BDSCA) explores the
applications of IoT in supply chain management and its benefits for organizations
and society. This review discusses the ethical, security, and operational
challenges of big data techniques with respect to IoT and privacy-preserving graph analytics.

There are numerous works focusing on IoT and privacy-preserving graph analytics in
the health domain. For example in ~\cite{srivastava2019}, the authors propose a
transactional protocol for remote patient monitoring using directed acyclic graphs.
This protocol is used to transfer patient data in a network of IoT wearable devices.
It uses a combination of a public and a private blockchain, and aims to resolve known
privacy and security issues for healthcare, without affecting scalability.
Another work~\cite{sharma2018} discusses the challenges in developing practical
privacy-preserving analytics in IoT-based healthcare information systems.
Parts of that discussion focus on existing privacy-preserving graph release methods.
Meeting these challenges --- including privacy-preserving graph data release ---
is critical for reliable healthcare IoT solutions.

\section{Empirical Studies and Open Research Questions}
\label{sec:discuss}

This Section describes existing open-source DP tools and discusses their limitations.
It also introduce a novel DP library that overcomes some of these limitations. It then
discusses possible research directions related to graph queries that are difficult to
make provably private.


\subsection{Implementations}
Despite the extensive literature devoted to graph differential privacy, there is a paucity of open source tools that implement these techniques. As such, it is frequently the case that researchers that want to use or extend these results will need to build their own tools to do so.  In many cases, a differentially private release mechanism for graph statistics can be decomposed into two pieces: a computation step that produces exact query responses on an underlying graph and a perturbation step that adds noise to the exact query responses to achieve some level of differential privacy.

Computing the exact query responses involves working with the underlying graph data structure. There are many open source tools which provide efficient implementations of common graph algorithms. See, for example,
\cite{Bromberger17},
\cite{csardi2006}, \cite{hagberg2008exploring}, and \cite{peixoto_graph-tool_2014}. In some cases, these algorithms can be used directly to compute the statistics of interest. In other cases, these algorithms are used to produce lower-sensitivity approximations to a query of interest. For example, in \cite{kasiviswanathan2013analyzing} network flows on graphs derived from the original graph are used to compute query responses on arbitrary graphs with sensitivity similar to that of the query restricted to graphs with bounded degree.  

In this setting, the perturbation step amounts to a straight-forward application of the appropriate differentially private release mechanism. While many such release mechanisms are mathematically simple, creating secure implementations is fraught with peril. As with cryptographic libraries, it is important to consider issues such as secure random number generation and robustness with respect to various side-channel attacks. Furthermore, because many differentially private release mechanisms operate on real numbers, as opposed to integers or floating point numbers that can be represented exactly in a computer, na{\"i}ve implementations of some mechanisms can be insecure. See for example \cite{mironov2012}. There are a number of open-source differential privacy libraries, see for example \cite{Gaboardi2020APF}, \cite{holohan2019diffprivlib}, \cite{rubinstein2017painfree}, and \cite{wilson2019differentially}. These libraries vary widely in the kinds of functionality that they provide, the security guarantees that they can offer, and the performance that they can achieve.  While most open-source libraries are sufficient for some tasks, e.g., research involving numerical simulations that are based on synthetic data, some may be insufficient for use in applications which will be used to protect real sensitive data.
 
We have produced reference implementations of some of the algorithms described in \cite{kasiviswanathan2013analyzing}\footnote{Available at: https://github.com/anusii/graph-dp} and \cite{nissim2007smooth}. 
These reference implementations use a new differential privacy library, RelM\footnote{Code and documentation are available at https://github.com/anusii/RelM}, developed by one of the authors to perturb the data prior to release. RelM provides secure implementations of many of the differentially private release mechanisms described in \cite{dwork2014algorithmic}.

In cases where the computation of differentially private query responses cannot be decomposed into separate computation and perturbation steps, the situation is grimmer.  We are unaware of any libraries that provide such functionality and as such any efforts to use or extend these algorithms will require the use of bespoke tools. 

\subsection{Empirical studies}

\mpp{
Many differentially private release mechanisms provide utility guarantees in the form of a bound on the probability that the difference between the perturbed and exact query responses will exceed some bound. These worst-case guarantees, however, do not necessarily describe how a release mechanism will perform when applied to a given data set. As such, many authors provide the results of experiments run on example data sets to demonstrate the average-case behavior of a proposed release mechanism.}

\mpp{
While many data sets have been used in such experiments, several have been used frequently enough to comprise a \emph{de facto} standard corpus. In particular, the data sets provided by the Stanford Network Analysis Project (SNAP) \cite{snapnets} have been used by multiple authors to test release mechanisms intended for use with a wide variety of graph analytics. Of these, data sets describing the collaboration network for papers submitted to various categories of the e-print arXiv (ca-HepPh, ca-HepTh, and ca-GrQc), the Enron email communication network (email-Enron), ego networks from various social networks (ego-Facebook, ego-Twitter, com-LiveJournal), and voting data for the election of Wikipedia administrators (wiki-Vote) were particularly popular.
}

\mpp{
In addition, many authors describe the performance of proposed release mechanisms on random graphs. The most common model used to generate such graphs in the papers we surveyed was the Erd\"os-R\'enyi-Gilbert (ERG) model. For mechanisms intended for use with scale-free networks,  the Barab\'asi-Albert model was frequently used to generate suitable random graphs.
} 


\kl{A recent work in \cite{xia2021dpgraph} has benchmarked selected edge DP and node DP mechanisms for privately releasing answers of degree sequence and subgraph counting queries. This work is implemented as a web-based platform \textit{DPGraph} with built-in DP-based graph mechanisms, real graph data that frequently appear in related empirical studies and visual evaluations to assist a user in selecting the appropriate mechanism and its privacy parameter for a project in hand. 
}

\kl{The empirical studies carried out in \cite{xia2021dpgraph} and \cite{ningbenchmarking} cover a wide range of DP-based graph mechanisms that have been reviewed in this survey paper, including
edge DP degree sequence~\cite{hay2009accurate},
node DP degree sequence~\cite{day2016publishing,kasiviswanathan2013analyzing,raskhodnikova2016lipschitz}, and
edge DP subgraph counting~\cite{chen2013recursive,karwa2011private,nissim2007smooth,zhang2015private}. 
}

\kl{The accuracy of these mechanisms is compared across a range of privacy budget values and their running time is compared across networks of different sizes. In general, graph size and shape, query type all have an impact on the performance of a mechanism, so the question of which mechanism to use with what parameter values is related to a specific project. Having said that, the authors also make some noticeable summaries of the different mechanisms. For private degree sequence release, \cite{hay2009accurate} has the best utility and running time under edge DP, while \cite{day2016publishing} has the best performance in both aspects for node DP. The performance of subgraph counting query mechanisms is more related to subgraph types. For example, \cite{karwa2011private} has the best performance for k-star counting, whilst \cite{zhang2015private} shows the lowest error for other subgraphs.}

\subsection{Useful but Difficult Graph Statistics}

Most of the graph statistics reviewed in \Cref{sec:stagraph} are related to subgraph queries and degree distribution queries. The former includes some common subgraphs such as triangle, $k$-triangle and $k$-star. The latter includes degree distribution/sequence and joint degree distribution. 
There are, however, many other graph statistics that are useful for understanding networks. One such class of metrics are centrality metrics. There are different types of centrality metrics, all of which measure the \emph{importance} or \emph{influence} of a node on the rest of the network. For example, degree centrality (equivalently node degree) measures how influential a node could be by looking at its direct neighbors. 
Betweenness centrality measures the importance of a node by counting the number of shortest paths it appears in between all pairs of nodes. 
Another important class of metrics are connectivity metrics.
For example, assortativity measures how well nodes of similar types (e.g., degrees)  connect to each other. 
Transitivity reflects to what extent edges in a path are transitive, e.g. if $x$ is a friend of $y$ who is a friend of $z$, how likely is it for $x$ and $z$ to be friends?  

Some of these metrics are high-level statistics, meaning they are calculated based on lower level statistics such as degree, neighbor or shortest path. Hence, they are more difficult to be made private while still remaining useful due to their high global sensitivity. \kl{A recent analysis in \cite{laeuchli2022analysis} concludes that the eigenvector, Laplacian and closeness centralities are extremely difficult to be made private using the smooth sensitivity technique.}
Research progress in private versions of such queries will significantly expand the application areas of private graph release mechanisms.

\section{Conclusion}\label{sec:end}


This paper provided a thorough survey of published methods for the provable
privacy-preserving release of graph data. It proposes a taxonomy to organize
the existing contributions related to private graph data, and use that taxonomy as a structure
to provide comprehensive descriptions of existing mechanisms, analytics, and their
application domains. At the top level, this survey differentiated between private
query release mechanisms and private graph release mechanisms. Within each category, we discussed existing non-provable and provable mechanisms, with significant emphasis on the latter type. Such provably
private mechanisms offer mathematical guarantees on formally defined privacy properties.
In contrast, non-provable methods lack such strong theoretical assurance. Most
provable techniques have been based on the concept of Differential Privacy (DP),
which we briefly introduced in the context of graph data.

For private graph statistics release, this survey further explored different
classes of methods, such as node DP, edge DP, local DP, with high-level descriptions
of numerous existing works. Similarly, we described several families of DP-based methods for the private release of synthetic graphs, such as generative graph models,
graph matrix perturbations, and iterative refinement. We then elucidated the
limitations of DP as pertinent to graph data release, and discussed several alternatives to DP that provide provable privacy, such as Pufferfish privacy (and the related Wasserstein mechanism),
Dependent DP, Adversarial Privacy, and Zero-Knowledge Privacy. We followed with 
a short review of several key application domains where the availability and use of
private graph data analytics is critical (e.g., as social networks, financial
service, supply chains, or health service sectors), with an emphasis on 
specific use cases within each domain. 
Finally, this survey concluded on some open issues in the field of private graph
analytics, such as the paucity of open source tools that implement the extensive
amount of surveyed mechanisms, and the fact that a wide range of useful graph
statistics do not have (yet) provably private counterparts (e.g., degree centrality),
thus providing potential leads for future research in the area.
This survey paper should benefit practitioners and researchers alike in the
increasingly important area of private graph data release and analysis.

\bibliographystyle{ACM-Reference-Format}
\bibliography{sample-base}

\end{document}